\documentclass[aps,prx,twocolumn,superscriptaddress,export,nofootinbib]{revtex4-1}
\usepackage[colorlinks=true,linkcolor=blue,citecolor=blue,urlcolor=blue]{hyperref}
\usepackage[utf8]{inputenc}
\usepackage[german, english]{babel}
\usepackage[dvipsnames]{xcolor}
\usepackage{amsmath,amsfonts, amssymb, amsthm, dsfont}
\usepackage{bm}
\usepackage{graphicx}
\usepackage{tikz}
\usepackage{physics}
\usepackage{xspace}
\usepackage[export]{adjustbox}

\usepackage{enumerate}
\usepackage{subcaption}
\usepackage[justification=raggedright,format=plain]{caption}
\usepackage{comment}

\newcommand{\sgn}{\operatorname{sgn}}

\newcommand{\Z}{\mathbb{Z}}

\newcommand{\be}{\begin{equation}}
\newcommand{\ee}{\end{equation}}
\newcommand{\ba}{\begin{aligned}}
\newcommand{\ea}{\end{aligned}}
\newcommand{\ri}{\mathrm{i}}

\graphicspath{{Figs/}}

\begin{document}

\title{
Yang-Lee Criticality as a Dissipative Dynamical Phase Transition: Quantum Simulation of non-Hermitian Physics without Post-selection
}
\author{Stephen W. Yan}
\affiliation{Department of Physics, University of California,
Santa Barbara, CA 93106, USA}
\date{\today}

\begin{abstract}
We show that the $d+0$-dimensional Yang-Lee theory describing classical Ising spins in an imaginary magnetic field can be realized, without post-selection, within a $(d-1)+1$ open quantum system whose dynamics consist of local unitaries and engineered dissipation.
Competition between the coherent unitary and dissipative dynamics drives a transition wherein the time-dependence of a particular class of linear observables changes from damped oscillatory ``underdamped'' to purely exponential ``overdamped'' decay.
Our construction relies on an extensive number of weak-symmetries of the Lindbladian fixed by the choice of observable but is otherwise exact.
Consequently, we show that the dynamics are described by the non-Hermitian generator of the Yang-Lee transfer matrix, leading to an effective Yang-Lee theory defined on the spacetime history of the open system.
By locally modifying the dynamics, we directly measure spin correlation functions of the Yang-Lee theory as well as a related ``Loschmidt Echo'' correlator which we detail.
We explicitly show that the required dissipation channels can be obtained through local $2$-qubit gates and discuss potential realization on near-term quantum simulator devices.
Finally, we generalize our construction to embed arbitrary non-Hermitian Hamiltonians within an open quantum system under Lindbladian time-evolution without post-selection, drawing connections between unconditional open quantum dynamics, exceptional point physics, and non-unitary statistical mechanics.
\end{abstract}

\maketitle
\tableofcontents

\section{Introduction}

Simulation in open quantum systems leverages an expanded toolkit of measurements and dissipative evolution, opening the door to a rich world of new quantum phenomena~\cite{Verstraete_2008,de_Groot_2022,Ma_2023,Diehl_2008,Ma_2025,Fan_2024,Wang_2025,skinner_2019,Li_2018,Chan_2019}.
For instance, measurement back-action, possibly followed by feedback or classical post-processing, can restructure or prepare interesting critical states~\cite{Lee_2023,Lee_2022,Lu_2023,Garratt_2023}.
In many cases, following measurement trajectories leads to randomness with intrinsic Born-rule correlations, altering the universality of critical points and potentially destabilizing entire phases~\cite{P_tz_2025,Zhu_2023,Jian_2023,Wang_2025_self_dual,Patil_2024,Yan_2026}.
In some cases, the novel criticalities are described by non-unitary conformal field theories for which analytic tools are lacking.

By contrast, certain classes of non-unitary critical points are well-understood, such as those corresponding to minimal models in $2+0$-dimensions~\cite{belavin_minimal,friedan_minimal}.  They serve as toy models for exploring the general consequences of non-unitarity.
A paradigmatic example is the Yang-Lee theory which has historically played an important role in understanding first-order phase transitions with various global symmetries~\cite{Yang_Lee_YL_Theory_I,Yang_Lee_YL_Theory_II,kurtze_yl_spherical,fisher_yl_1d}.
The Yang-Lee theory can also be viewed as a classical statistical mechanical model in its own right, microscopically realized with Ising spins
\begin{align}\label{eqn:yl_classical_partition_function}
    \mathcal{Z}_\mathrm{YL} &= \sum_{\{\sigma\}}
    e^{- \beta H_\mathrm{YL}[\{\sigma\}]}
    \\
    H_\mathrm{YL}[\{\sigma\}] &= -J \sum_{\langle i , j\rangle} 
    \sigma_i \sigma_j - \ri h \sum_{i} \sigma_i
    \,,
\end{align}
in the presence of an external, albeit imaginary, field~\cite{Fisher_YL_Theory}.
Criticality, also known as the ``edge singularity'' occurs in the paramagnetic $T > T_c$ phase of the model as the field is tuned to a critical value $h_c \sim (T-T_c)^{15/8}$~\cite{fonseca_yl_scaling,mangazeev_yl_critical_location}.
The critical theory enjoys conformal invariance and is described by the minimal model $\mathcal{M}_{2, 5}$ whose non-unitarity allows for striking features such as negative scaling dimensions and negative central charge~\cite{Cardy_YL_Theory}.

Given the analytic control over such theories, 
it is natural to ask if they may be realized as critical points of physical systems.
However, direct realization of Yang-Lee theory is complicated due to the imaginary field.
Consequently, early proposals indirectly studied Yang-Lee by analytically continuing measurements made for real fields~\cite{binek_anal_cont,deger_fluctuations}.
Recently, additional proposals have been put forth which simulate the Yang-Lee problem more directly, 
but these have either relied on non-local couplings to a probe spin~\cite{wei_central_decoherence_proposal,Wei_BB_Central_Spin, probing_central_charge,experimental_peng,Francis_Central_Spin}, non-Hermitian time-evolution through post-selection~\cite{Matsumoto_YL_exp,krishnan_2019} or systems whose dynamics are only approximately non-Hermitian~\cite{Gao_YL_exp}.

In this work, we study open quantum systems evolving under competing unitary and dissipative dynamics.
Here, Yang-Lee theory naturally arises as
describing a certain class of observables linear in the density matrix for which the Lindbladian maps exactly to the non-Hermitian  generator $\hat{H}_\mathrm{YL}$ of the Yang-Lee transfer matrix.
The required dissipation 
in the unconditional dynamics results from either averaging over many measurement trajectories with feedback or by coupling to ancillae degrees of freedom which are subsequently ``traced out'' and discarded~\cite{harrington_engineered_dissipation}.
We propose a concrete scheme for realizing our proposal in terms of simple measurements and local gates coupling to the ancillae which is achievable on near-term quantum simulators.
Consequently, we may tune the relative strengths of the unitary and dissipative dynamics, providing access to the Yang-Lee critical point, where we also devise a protocol for directly measuring spin correlation functions and a related ``Loschmidt Echo'' correlator through local modifications of the dynamics.

Our results show that Yang-Lee theory is a natural description of the spacetime dynamics of an open quantum system when probed by particular observables.
Furthermore, the Yang-Lee critical point reached by tuning the imaginary field strength is understood as the point where the time-evolution of these observables undergoes a sharp transition between oscillatory ``underdamped'' and exponential ``overdamped'' decay.
In turn, the same physics is described by the exceptional point of $\hat{H}_\mathrm{YL}$ where the $\mathcal{PT}$-symmetry is broken which itself is related to the $\mathbb{Z}_2$ Hermiticity-preserving symmetry of the Lindbladian of the open quantum system.

The construction relies on an extensive number of weak-symmetries of the Lindbladian leading to a ``fragmentation'' of the Hilbert space~\cite{essler2020integrability,
PhysRevResearch.5.043239,PhysRevB.109.054311} while the choice of observable projects into a \textit{non-Hermitian subspace} in which the Lindbladian is equivalent to $\hat{H}_\mathrm{YL}$.
We are able to show that this structure is generic.
That is, given a generic non-Hermitian Hamiltonian in any dimension, we show that one can always construct a parent Lindbladian with an extensive number of weak-symmetries such that the time-evolution of particular observables is determined by the target Hamiltonian.
This can be done while preserving locality and requires only a constant number of ancillae per system qubit.
Our result shows that any non-Hermitian Hamiltonian can be realized from Lindbladian time-evolution without post-selection~\cite{Abo_YL_exp,Minganti_NHH_no_go}.

The rest of the paper will be organized as follows.
In Sec.~\ref{sec:non_hermitian_primer} we formulate the $d+0$-dimensional Yang-Lee theory in terms of non-Hermitian time-evolution in $(d-1)+1$-dimensions and outline the general strategy for its realization through weakly-symmetric Lindbladians.
In Sec.~\ref{sec:yang_lee_construction}, we detail our construction for the Yang-Lee theory as a discrete-time dynamics realizable on digital quantum simulators.
We also discuss the direct measurement of various spin correlation functions in our framework.
We extend our construction in Sec.~\ref{sec:general_nhh_without_postselection} to realize generic non-Hermitian Hamiltonians in a post-selection free way.
Finally, we conclude in Sec.~\ref{sec:conclusion} with a discussion of possible experimental realization and potential future directions.

\section{Non-Hermitian Transfer Matrices from Dissipation Engineering}\label{sec:non_hermitian_primer}
\begin{figure*}[t]
    \centering
    \includegraphics[width=.95 \textwidth  ]{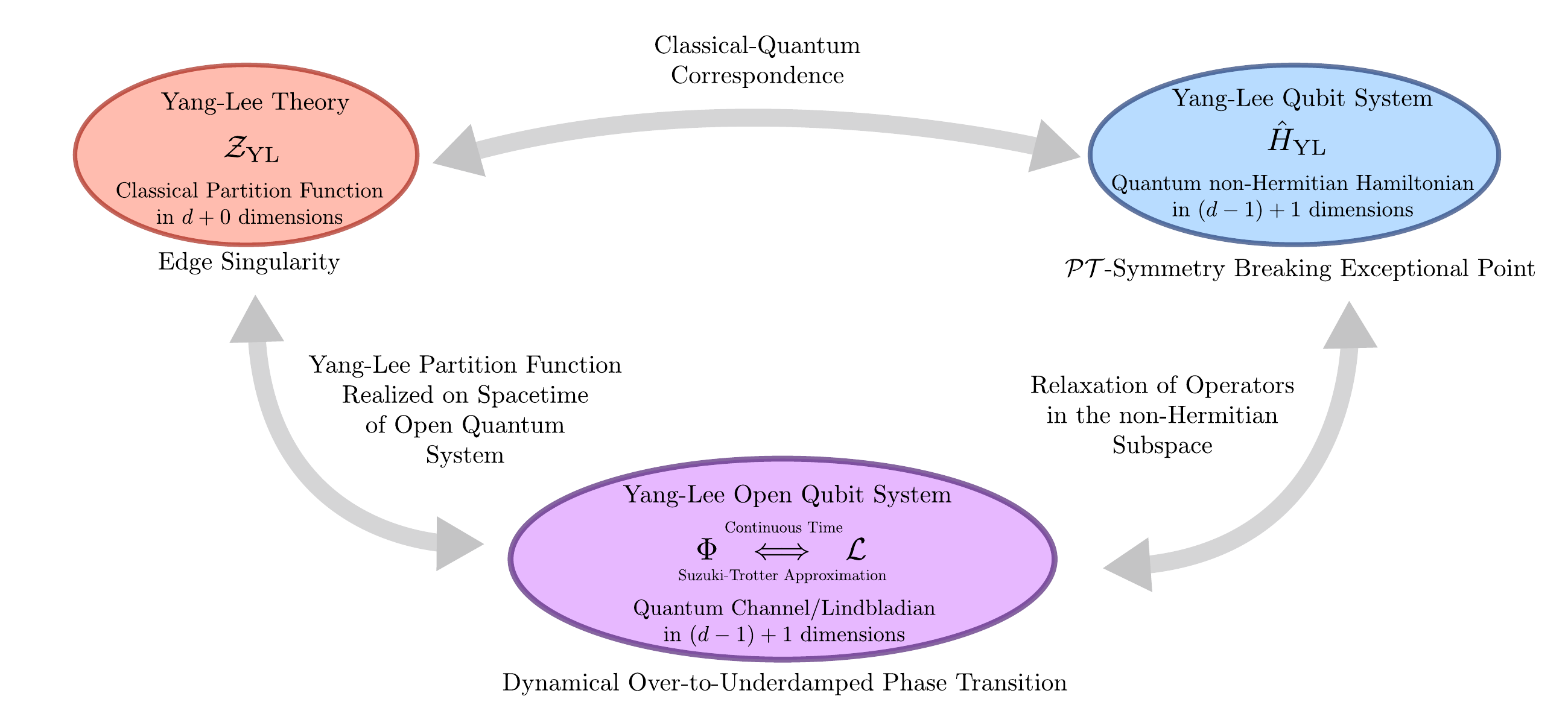}
    \caption{Overview of the main ideas of this paper allowing us to realize Yang-Lee theory in an open quantum system without the use of post-selection.
    The classical partition function $\mathcal{Z}_\mathrm{YL}$ is related to $\hat{H}_\mathrm{YL}$ through the standard classical-quantum mapping which in turn emerges as the description of the decay of operators in the non-Hermitian subspace of an open quantum system.}\label{fig:lee_yang_mappings}
\end{figure*}
In this section, we outline the general strategy for embedding non-Hermitian physics within the Lindbladian time-evolution of particular operators, summarized in Figure~\ref{fig:lee_yang_mappings}.
We center our discussion on the application to Yang-Lee theory, but the same strategy is generalized to other non-Hermitian Hamiltonians in Sec.~\ref{sec:general_nhh_without_postselection}.

To begin, we relate the $d+0$-dimensional classical Yang-Lee statistical mechanical model of Ising spins in an imaginary field~\eqref{eqn:yl_classical_partition_function} to a $(d-1)+1$-dimensional dynamics of an open quantum system through the standard ``classical-to-quantum'' correspondence~\cite{Fradkin_field_theory}.
As a review, the mapping begins by choosing a ``quantization'' axis $t$ of the classical theory with the remaining $d-1$ orthogonal axes retained as spatial directions.
The partition function can then be written in terms of transfer matrices $\hat{T}$ acting between spatial configurations at adjacent times
\begin{equation}
    \mathcal{Z}_\mathrm{YL} = \bra{\phi_f} \left(\hat{T}\right)^T \ket{\phi_i} \,,
\end{equation}
for some boundary states $\ket{\phi_{i, f}}$ and where $T$ is the length of the system in the time direction.
Next, the model is made anisotropic by allowing the Ising coupling $\beta J_{t}$ and $\beta J_{x}$ to vary in the time and spatial dimensions, respectively.
In this way, we pass to the continuous-time limit via 
\begin{align}
    \mathcal{Z}_\mathrm{YL} &=  \lim_{\delta t \to 0} \bra{\phi_f} 
    \left(\hat{T}[\delta t]\right)^{T/\delta t}
    \ket{\phi_i}
    \\
    &= \bra{\phi_f} e^{-T \hat{H}_\mathrm{YL}} \ket{\phi_i}
    \,,\label{eqn:yl_Z_as_transfer_matrix}
\end{align}
where $\hat{T}[\delta t] \approx 1 - \delta t \hat{H}_\mathrm{YL}$ is the infinitesimal transfer matrix obtained by sending 
$\beta H \sim \beta J_x \sim e^{-\beta J_t} \sim \delta t \to 0$ while keeping the ratios finite.

The partition function is thus mapped to the imaginary-time evolution
of the quantum non-Hermitian Hamiltonian
\begin{equation}\label{eqn:yl_nhh_form}
    \hat{H}_\mathrm{YL} = 
    -\sum_{\langle i, j\rangle}
    \gamma \hat{\tau}^z_{i} \hat{\tau}^z_{j}
    - \sum_{i} 
    \Delta \hat{\tau}^x_{i} + \ri \theta \hat{\tau}^z_i
    \,,
\end{equation}
where $\hat{\tau}^{x, z}$ are Pauli operators acting on the Hilbert space spanned by basis states corresponding to Ising spin configurations on the $(d-1)$-dimensional spatial time-slice.
This Hamiltonian is symmetric under conjugation by the $\mathcal{PT}$-symmetry generator $\mathcal{K} \prod_i \hat{\tau}^x_i$ where $\mathcal{K}$ denotes complex conjugation~\cite{Bender_PT_Sym,Mostafazadeh_PT_Sym}.
The $\mathcal{PT}$-symmetry of the quantum transfer matrix is inherited from the $\mathcal{PT}$-symmetry of the classical model acting on Ising spins as $\sigma \to - \sigma$ followed by complex conjugation.
Furthermore, the Yang-Lee critical point is identified with a $\mathcal{PT}$-symmetry breaking exceptional point of the non-Hermitian Hamiltonian~\cite{gehlen_yl_as_nhh}.
Criticality is obtained by tuning the dimensionless field $h := \theta / \Delta$ to a critical value $h \to h_c(\gamma / \Delta)$ within the paramagnetic phase $\gamma / \Delta < 1$.

\subsection{Operator Dynamics in Doubled Hilbert Space}\label{sec:operator_dyn_doubled_H}
The non-Hermitian dynamics under the generator of the Yang-Lee transfer matrix~\eqref{eqn:yl_nhh_form} can be embedded, in a sense to be made precise shortly, within the quantum dynamics of an open system of qubits evolving continuously under the following Lindbladian
\begin{align}\label{eqn:yl_lindbladian_form}
    \mathcal{L}[\rho] = 
    \ri \frac{\theta}{2}
    \sum_j
    [Z_j,& \rho]
    + \Delta \sum_j X_j \rho X_j 
    \\
    &+ \gamma\sum_{\langle i, j\rangle} 
    e^{-\ri \frac{\pi}{4} (Z_i-Z_j)}
    \rho
    e^{\ri \frac{\pi}{4} (Z_i-Z_j)}
    \,.
    \nonumber
\end{align}
We have dropped a constant term fixing the overall normalization of the state.
Here, $X, Z$ denote the Pauli matrices acting on the open qubit system and $\theta, \Delta, \gamma$ are parameters of the dynamics which can be freely tuned to realize the Yang-Lee critical point.
We justify the form of the Lindbladian and show how each term can be concretely realized on a quantum simulator device in Sec.~\ref{sec:yang_lee_construction}.

The Lindbladian superoperator generates the continuous time-evolution of the density matrix~\cite{lindblad_dyn_semigroup,gorini_dyn_semigroup}
\begin{equation}
    \frac{d \rho}{dt} = \mathcal{L}[\rho] \implies 
    \rho(T) = e^{T \mathcal{L}}[\rho(0)] \,.
\end{equation}
To elucidate the connection with a non-Hermitian Hamiltonian acting on pure states, we use the Choi-Jamio\l kowski isomorphism
to relate density matrices $\rho \in L(\mathcal{H})$ in the space of linear operators to pure states in the ``doubled Hilbert space'' $\mathcal{H} \otimes \mathcal{H}^*$.
States in the doubled Hilbert space are denoted $| \rho \rangle\rangle$, with the isomorphism sending $\rho_{ij} \ket{i}\bra{j} \to \rho_{ij} \ket{i} \otimes \ket{j}$
where $\otimes$ is the tensor product between the two copies.
The Lindbladian is then realized as a linear operator acting on states in the doubled Hilbert space and takes the form
\begin{align}
    \mathcal{L}
    = 
    \ri \frac{\theta}{2}
    \sum_j
    Z_j &\otimes I_j - I_j \otimes Z_j
    + \Delta \sum_j X_j\otimes X_j 
    \\
    &+ \gamma\sum_{\langle i, j\rangle} 
    e^{-\ri \frac{\pi}{4} (Z_i-Z_j)}
    \otimes
    e^{\ri \frac{\pi}{4} (Z_i-Z_j)}
    \,.
    \nonumber
\end{align}

Additionally, we are interested in linear observables of the time-evolved density matrix, which map to scattering amplitudes in the doubled Hilbert space.
First, the time-evolved density matrix $|\rho(T) \rangle \rangle = e^{T \mathcal{L}} | \rho(0)\rangle \rangle$ is obtained by evolving an initial state $| \rho(0) \rangle \rangle$ determined by the initial density matrix $\rho(0)$.
Expectation values of the density matrix are obtained by computing the overlap of $| \rho(T) \rangle\rangle$ with a particular reference state.
Let $\ket{i}$ denote a basis of $\mathcal{H}$ with respect to which the operator $O$ is diagonal and $O_i$ the diagonal matrix elements in that basis.
Then the appropriate reference state takes the form $\langle \langle O | = \sum_i O_i \bra{i} \otimes \bra{i}$.
The expectation value of $O$ is expressed as
\begin{equation}\label{eqn:equiv_O_exp_val_and_doubled_overlap}
    \Tr\left( O \rho(T) \right)= 
    \langle \langle O | e^{T \mathcal{L}} | \rho(0)\rangle \rangle \,,
\end{equation}
which can be further mapped to the representation of the Yang-Lee partition function in terms of the transfer matrix~\eqref{eqn:yl_Z_as_transfer_matrix}.
The initial and final states $\ket{\phi_i}, \bra{\phi_f}$ are obtained by appropriate choice of the initial density matrix and observable $|\rho(0)\rangle\rangle$, $\langle \langle O|$, respectively.
On the other hand, $\hat{H}_\mathrm{YL}$ is obtained 
by projecting into a subspace of $L(\mathcal{H})$ in which the Lindbladian is exactly described by the non-Hermitian Hamiltonian.
The projection is achieved through the choice of observable $O$.
The details follow below.

\subsection{Extensive Weak-Symmetry and the non-Hermitian Subspace}
Consider the linear subspace of operators
\begin{equation}\label{eqn:nhh_subspace}
    \Lambda := \bigotimes_i
    \mathrm{span}_{\mathbb{C}}\left\{X_i, Y_i \right\}
\end{equation}
spanned by Pauli strings consisting of $X$ or $Y$ on each site.
We call $\Lambda$ the \textit{non-Hermitian subspace}.
Observe that $\Lambda$ is invariant under the time-evolution generated by the Lindbladian superoperator.
Consequently, when the observable $O$ is chosen to lie within the non-Hermitian subspace $\Lambda$, its expectation value in Eq.~\eqref{eqn:equiv_O_exp_val_and_doubled_overlap} is completely characterized by the projection of $\mathcal{L}$ into $\Lambda$.

Indeed, the dynamics restricted to $\Lambda$ is isomorphic to that of a qubit system under non-Hermitian Hamiltonian time-evolution.
In particular, for each physical site we associate the operators $X_i$ and $\ri Y_i$ with the qubit states $\ket{+}_i, \ket{-}_i$, respectively.
Formally, we have the identification
\begin{align}\label{eqn:nhh_projection_dictionary}
    \mathcal{Z}_\mathrm{YL} = \bra{\phi_f} e^{-T \hat{H}_\mathrm{YL}} &\ket{\phi_i} = \Tr\left( O e^{T \mathcal{L}}[\rho(0)]\right) \\
    \hat{H}_\mathrm{YL} &\cong -P_\Lambda \mathcal{L} P_\Lambda
    \\
    \ket{\phi_i} &\cong P_\Lambda | \rho(0) \rangle\rangle \\
    \ket{\phi_f} &\cong P_\Lambda | O \rangle\rangle
    \,,
\end{align}
with $P_\Lambda$ the projector into $\Lambda$.
For generic $\mathcal{L}$, the projection into $\Lambda$ results in a non-Hermitian Hamiltonian.
We show in Sec.~\ref{sec:yang_lee_construction} that the Lindbladian~\eqref{eqn:yl_lindbladian_form} reproduces the desired non-Hermitian Hamiltonian~\eqref{eqn:yl_nhh_form} that generates the Yang-Lee transfer matrix.
Thus, by measuring the observable $O$, we realize the Yang-Lee partition function defined on the spacetime history of the open quantum system.

In order for the expectation value of $O$ to be non-trivial, the initial state $\rho(0)$ must be chosen to have finite overlap with the non-Hermitian subspace.\footnote{Note that it is the choice of the observable $O$ which projects us into the non-Hermitian subspace.  
The initial state $\rho(0)$ cannot be chosen to lie completely within $\Lambda$ because physical states satisfy $\Tr \rho(0) = 1$.}
This has the following physical interpretation--the maximally-mixed state is the steady state of the Lindbladian in which all observables have trivial expectation value.
Thus, $\Tr \left( O \rho(T) \right)$ can only describe the relaxation from an initial state in which $O$ has non-trivial expectation value.
For concreteness, we consider $O = \prod_i X_i$ and the initial state $\rho(0) = \prod_i \ket{+}_i \bra{+}_i$ which has maximal expectation value $+1$ for $O$.
Other choices of $O$ and $\rho(0)$ modify the boundary conditions of $\mathcal{Z}_\mathrm{YL}$ but otherwise do not change the universal features of the bulk physics, as long as $O \in \Lambda$ and $\rho(0)$ has finite overlap with $\Lambda$.

\begin{figure*}[t]
    \centering
    \includegraphics[width=.99 \textwidth  ]{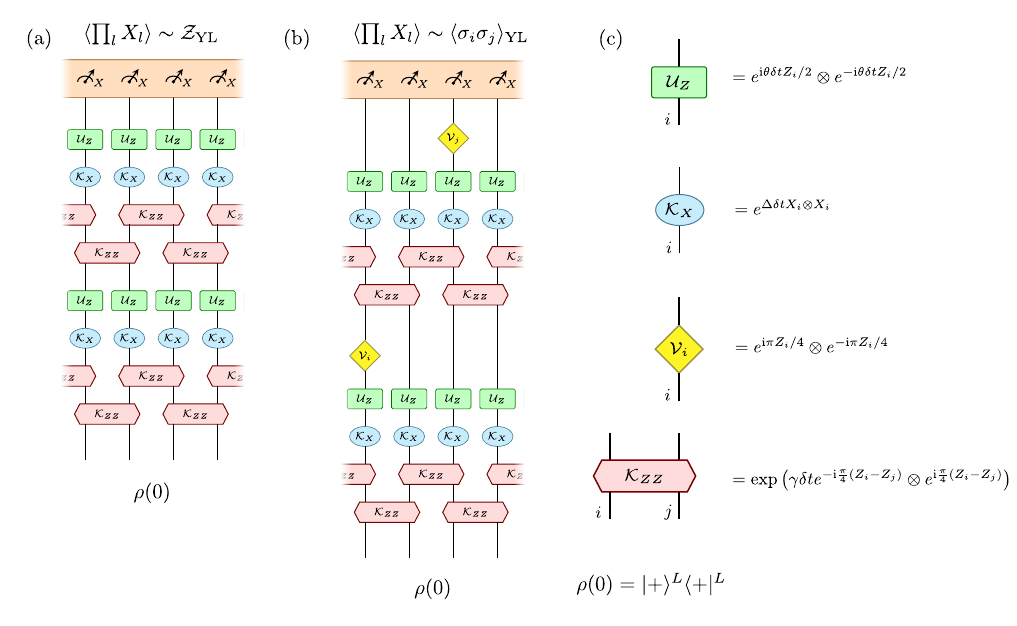}
    \caption{
    In (a), we diagram the quantum circuit which realizes $\hat{H}_\mathrm{YL}$ and consequently, the Yang-Lee theory with engineered dissipation but without post-selection.
    At the final time-step, we measure $\prod_j X_j$, whose expectation value is exactly identical to the classical Yang-Lee partition function defined on spacetime.
    In (b), the circuit is modified by single-site unitaries $\mathcal{V}$ whose insertion allows us to extract spin-spin correlation functions by measuring $\prod_j X_j$.
    In (c), we diagram the circuit elements and corresponding initial state.
    A concrete way to realize the two-site decoherence channel $\mathcal{K}_{ZZ}$ is provided in Appendix~\ref{sec:app_circ_realization}.
    }\label{fig:circit_diagram}
\end{figure*}

More generally, the non-Hermitian dynamics arise as an effective description of the relaxation process of observables $O \in \Lambda$ rather than as a property of the Lindbladian steady state.
Indeed, it is known from general arguments that the steady-state properties cannot exhibit exceptional point physics present for non-Hermitian Hamiltonians~\cite{Minganti_NHH_no_go}.
In fact, our non-Hermitian subspace corresponds to an excited-state manifold of the original Lindbladian and was known to the authors of Ref.~\cite{Minganti_NHH_no_go}.
Our present work differs in that we demonstrate a general construction to realize any non-Hermitian exceptional point within open quantum dynamics without post-selection in Sec.~\ref{sec:general_nhh_without_postselection}, with application to the Yang-Lee critical point in Sec.~\ref{sec:yang_lee_construction}.

Before proceeding, we comment that the existence of the non-Hermitian subspace $\Lambda$ can also be understood as arising from an extensive number of weak-symmetries~\cite{Buca_Prosen_Weak_Symmetry,Albert_Jiang_Weak_Symmetry} of the Lindbladian of the form $Z_i \otimes Z_i$ for each site $i$.
The subspace $\Lambda$ is precisely the subspace of $L(\mathcal{H})$ with eigenvalue $-1$ of the weak-symmetry.
A similar construction was used to reduce the Lindbladian dynamics in an unrelated system in Ref.~\cite{yan_ctfim} to a non-Hermitian Hamiltonian describing the complex-valued transverse-field Ising model and served as initial motivation for this work.
For completeness, we formulate our result in the language of weak-symmetries in Appendix~\ref{sec:app_weak_gauge_sym}.

\section{Yang-Lee Criticality without Post-selection}\label{sec:yang_lee_construction}
In this section, we realize Yang-Lee theory~\eqref{eqn:yl_nhh_form} as the effective non-Hermitian Hamiltonian describing the relaxation dynamics of operators in the non-Hermitian subspace $\Lambda$.
Our mapping between the Lindbladian~\eqref{eqn:yl_lindbladian_form} and the desired non-Hermitian Hamiltonian~\eqref{eqn:yl_nhh_form} is formally exact and does not rely on post-selected dynamics.
For simplicitly, we focus on the $2+0$-dimensional Yang-Lee theory, which can be realized via the open quantum dynamics of a $1+1$-dimensional spin chain.
The construction is easily generalized to arbitrary dimensionality.

Additionally, in place of directly implementing the continuous-time evolution described by $\mathcal{L}$, we provide a detailed construction of a discrete-time approximation which is more applicable for digital quantum simulator platforms.
The discrete-time dynamics are defined by the Suzuki-Trotter decomposition of~\eqref{eqn:yl_lindbladian_form} and reproduce the Lindbladian time-evolution operator $e^{T\mathcal{L}}$ in the continuous-time limit $\delta t \to 0$.
For finite $\delta t$, we incur Trotter errors on the order $\mathcal{O}(\delta t^2)$, but these can be suppressed by decreasing $\delta t$.
We numerically simulate the discrete-time evolution and verify that the universal features of the Yang-Lee critical point are qualitatively unaffected.
\subsection{Setup}\label{sec:yl_setup}
We consider a length $L$ qubit chain under dynamics defined by the quantum channel
\begin{align}
    \Phi &= \underbrace{\Phi[\delta t] \circ \Phi[\delta t] \circ \cdots \circ \Phi[\delta t]}_{T/\delta t}
    \\
    \Phi[\delta t] &= 
    \mathcal{U}_Z[\delta t] \circ \mathcal{K}_X[\delta t] \circ \mathcal{K}_{ZZ}[\delta t] \,.
    \label{eqn:single_time_step_channel}
\end{align}
which is the Suzuki-Trotter decomposition~\cite{trotter,suzuki} of~\eqref{eqn:yl_lindbladian_form}.
Here $\mathcal{U}_Z[\delta t]$, $\mathcal{K}_X[\delta t]$ and $\mathcal{K}_{ZZ}[\delta t]$ are a single-site unitary, single-site $X$-dephasing channel and a particular nearest-neighbor channel parameterized by strengths $\theta \delta t$, $\Delta \delta t$ and $\gamma \delta t$, respectively, whose precise forms we discuss shortly.
We justify the form of each component by examining their action on the operators $X$, $\ri Y$ which form a convenient basis for $\Lambda$.

The channel after two time-steps is pictured in Figure~\ref{fig:circit_diagram}a.

\subsubsection{Single-Site \texorpdfstring{$X$}{X}-Dephasing Channel}
We consider the single-site $X$-dephasing channel.
The channel can be written
\begin{equation}
    \mathcal{K}_X = \prod_{j=1}^L \mathcal{K}_{X, j}
    \,,
\end{equation}
where each $\mathcal{K}_{X, j}$ acts within the 
doubled Hilbert space as
\begin{equation}
    \mathcal{K}_{X, j} = 
    (1-\Delta \delta t ) I_j \otimes I_j + \Delta \delta t X_j \otimes X_j \sim
    e^{\Delta \delta t X_j \otimes X_j}
    \,,
\end{equation}
up to a constant normalization factor which we drop.
The $X$-dephasing channel can be engineered by weakly measuring the constituent qubit degrees of freedom in the $X$ basis with strength $\sqrt{\Delta \delta t}$ and averaging over many independent runs of the experiment without conditioning on measurement outcomes.

Next, observe that the channel acts by $X$ conjugation such that $X \to X$ while $\ri Y \to -\ri Y$ accrues a sign.
Extending this action to the entirety of the non-Hermitian subspace, we see that
\begin{equation}
\mathcal{K}_X \sim
    e^{\Delta \delta t \sum_{j} \hat{\tau}_j^x} \,,
\end{equation}
which is the second term in $\hat{H}_\mathrm{YL}$~\eqref{eqn:yl_nhh_form}.

\subsubsection{Single-Site Unitary}
Next, we consider single-site unitary rotations around the $Z$ axis about an angle $\theta \delta t / 2$.
The total rotation acting on all sites takes the form
\begin{equation}
    \mathcal{U}_Z
    = 
    \prod_{j=1}^L
    e^{\ri \theta \delta t Z_j / 2}
    \otimes 
    e^{-\ri \theta \delta t Z_j / 2}
    \,.
\end{equation}
For finite $\delta t$, the unitary acts on $L(\mathcal{H})$ by conjugation,
which becomes the adjoint action in the continuous-time limit.
Thus, the action of the unitary on an operator $\sigma_j$ supported at site $j$ takes the form
\begin{equation}
    e^{\ri \theta \delta t Z_j / 2} \sigma_j e^{- \ri \theta \delta t Z_j / 2} 
    \approx
    e^{\ri \theta \delta t [Z_j , \sigma_j] / 2}
    \,,
\end{equation}
Using the 
commutation relations
\begin{equation}
        [Z, X] = 2 \ri Y   \,,\;\;\;\;
        [Z, \ri Y] = 2 X
        \,,
\end{equation}
we conclude that
\begin{equation}
    \mathcal{U}_Z \sim
    e^{
    \ri  \theta \delta t
    \sum_j \hat{\tau}^z_j
    }
    \,,
\end{equation}
which is the third term in $\hat{H}_\mathrm{YL}$~\eqref{eqn:yl_nhh_form}.

\subsubsection{Nearest-Neighbor Channel}
Finally, the following nearest-neighbor channel generates the nearest-neighbor Ising interaction
\begin{equation}\label{eqn:form_of_the_NN_channel}
    \mathcal{K}_{ZZ} = \prod_{j=1}^L
    \exp \left(
    \gamma \delta t \, 
    e^{- \ri \frac{\pi}{4}(Z_j - Z_{j+1})}
    \otimes
    e^{\ri \frac{\pi}{4}(Z_j - Z_{j+1})}
    \right)
    \,.
\end{equation}
In the continuous-time limit, the channel can be unraveled in terms of the following Kraus operators up to overall normalization
\begin{align}\label{eqn:2site_lindbladian_operator}
\exp &\left(
    \gamma \delta t \, 
    e^{- \ri \frac{\pi}{4}(Z_j - Z_{j+1})}
    \otimes
    e^{\ri \frac{\pi}{4}(Z_j - Z_{j+1})}
    \right)
    \\
&\;\;\;\;\;\;\;\;\;\;\;\;\;\;\;\;\;\;\;\;\;\;\;\;\;\;\;\;\;\;\;\;\;\;\;\;\;\;\;
    \sim 
    \sum_{\mu_{j, j+1}} K_{\mu_{j, j+1}} \otimes K_{\mu_{j, j+1}}^*
    \nonumber
\\
        K_{\mu_{j, j+1}} &= 
    \frac{
    \exp
    \left[
    \mu_{j, j+1}
    \sqrt{\gamma \delta t} 
    (
    I + 
    Z_j Z_{j+1}
    - \ri Z_j
    + \ri Z_{j+1}
    )
    /2
    \right]
    }
    {
    \sqrt{
    2 \cosh 
    \left[
    \sqrt{\gamma \delta t}
    (I + Z_j Z_{j+1})
   \right] 
   }
    }
    \label{eqn:2site_kraus_operator}
    \,.
\end{align}
In Appendix~\ref{sec:app_circ_realization}, we provide a concrete circuit realization of this Kraus operator.
At a high level, it consists of a weak measurement of $Z_j Z_{j+1}$ followed by single-site phase rotation around the $Z_j$ and $Z_{j+1}$ axes about a rotation angle whose overall sign is conditioned on the measurement outcome $\mu_{j, j+1} = \pm 1$.
The feedback procedure does not necessarily have to be done in-time; instead of weak measurement, one can perform a controlled rotation onto an ancilla, followed by controlled phase rotations from the ancilla back into the system, with the ancilla being projectively measured at the end of the dynamics.
Finally, the unconditional quantum channel $\mathcal{K}_{ZZ}$ arises after averaging the measurement outcome $\mu$ over many runs of the experiment.

To see that $\mathcal{K}_{ZZ}$ maps to the desired nearest-neighbor Ising interaction within the non-Hermitian subspace, observe that 
\begin{equation}
    e^{-\ri \pi Z_j/4}
    X
    e^{\ri \pi Z_j/4} = Y
    \,,\;\;\;\;
        e^{-\ri \pi Z_j/4}
    Y
    e^{\ri \pi Z_j/4} = -X
    \,.
\end{equation}
Thus the argument of the exponential in Eq.~\eqref{eqn:form_of_the_NN_channel} takes $X_{j}X_{j+1} \leftrightarrow -Y_jY_{j+1}$ and $X_jY_{j+1} \leftrightarrow Y_jX_{j+1}$.
Since the non-Hermitian subspace is identified with the Hilbert space on which $\hat{H}_\mathrm{YL}$ acts through $X \mapsto |+\rangle$ and $\ri Y \mapsto |-\rangle$, we conclude that the action corresponds to the Ising interaction $\hat{\tau}^z_{i} \hat{\tau}^z_{i+1}$.
In other words,
\begin{equation}
    \mathcal{K}_{ZZ} \sim e^{\gamma \delta t \sum_j \hat{\tau}^z_j \hat{\tau}^z_{j+1}}
    \,,
\end{equation}
which is the first term in $\hat{H}_\mathrm{YL}$~\eqref{eqn:yl_nhh_form}.

\begin{figure*}[t]
    \centering
    \includegraphics[width=.99 \textwidth  ]{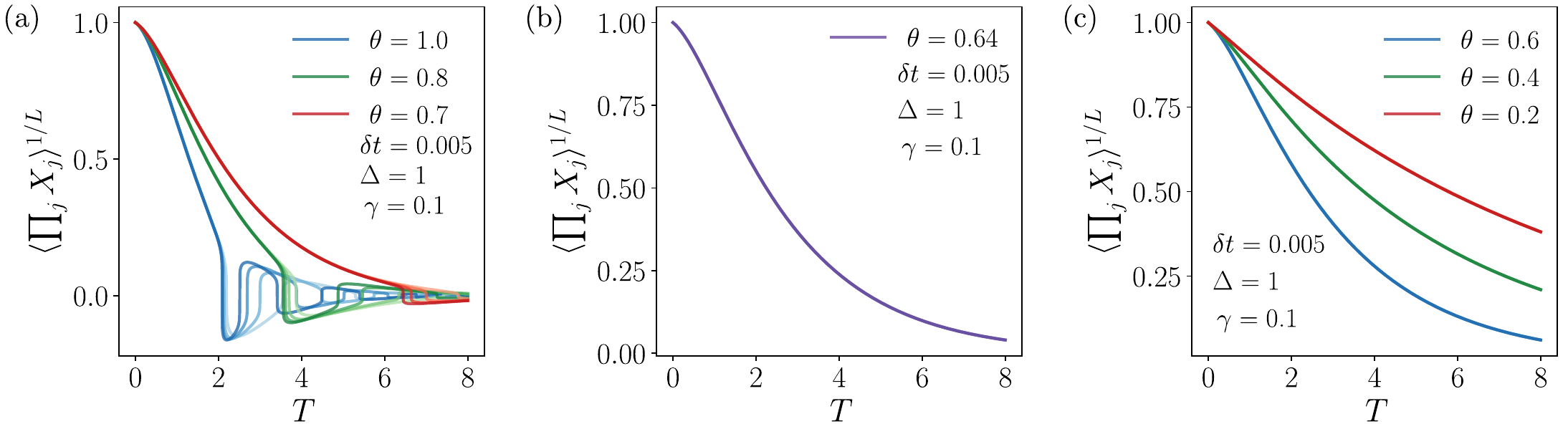}
    \caption{The expectation value $\langle \prod_j X_j \rangle \equiv \Tr \prod_j X_j \rho(T)$ evaluated at time $T$ starting from the initial state $\rho(0) = \ket{+}^{L}\bra{+}^L$.
    The $1/L$-th root is defined for negative arguments as $(\cdot)^{1/L} \equiv \sgn(\cdot) \times |\cdot|^{1/L}$.
    We demonstrate (a) the underdamped phase $\theta > \theta_c$, (b) the critically damped point $\theta = \theta_c$ and (c) the overdamped phase $\theta < \theta_c$.
    The over-to-underdamped transition coincides with the Yang-Lee critical point.
    The lighter to darker color indicates increasing system size $L=4,5, 6, 7$.
    In (b) and (c), the curves for different $L$ and the same $\theta$ lie on top of another.
    }\label{fig:over_to_underdamped}
\end{figure*}

\subsubsection{Choice of Observable and Initial State}
The choice of observable $O$ and the initial state $\rho(0)$ determine the boundary states $\bra{\phi_f}, \ket{\phi_i}$ of the transfer matrix and consequently the boundary conditions of the Yang-Lee partition function.

In order to project into the non-Hermitian subspace in which $\mathcal{L}$ reduces to evolution by $\hat{H}_\mathrm{YL}$, the observable $O$ must be chosen to lie within $\Lambda$.
For simplicity, we choose $O = \prod_j X_j$.  Under the correspondence $X \mapsto \ket{+}$, we see that $\bra{\phi_f} = \bra{+}^L$.

Similarly, the simplest initial state which has a non-trivial expectation value for this choice of $O$ is the state 
\begin{equation}
    \rho(0) = 
\prod_j \ket{+}_j \bra{+}_j
=
    \prod_j \left(
    \frac{1 + X_j}{2}
    \right)
    \,.
\end{equation}
It is clear that $\rho(0) \not\in \Lambda$.
Expanding the product, we see that the only term lying within $\Lambda$ is $\prod_j X_j$ which again maps to $\ket{+}^L$.
Thus, we conclude that
\begin{align}
    \Tr \left(\prod_j X_j e^{T \mathcal{L}}[\rho(0)] \right)
    &= \bra{+}^L e^{- T \hat{H}_\mathrm{YL}} \ket{+}^L 
    \\
    &= \mathcal{Z}^\mathrm{free}_\mathrm{YL}
    \,,
\end{align}
such that the expectation value of $O$ maps to to the Yang-Lee partition function with free boundary conditions.
\subsection{Yang-Lee Criticality as a Dynamical Phase Transition}

By mapping Yang-Lee theory~\eqref{eqn:yl_classical_partition_function} to the Lindbladian evolution~\eqref{eqn:yl_lindbladian_form} of an open quantum system, we identify the critical theory as the effective description of an ``over-to-underdamped'' transition of operators $O \in \Lambda$~\cite{yan_ctfim}.

The open quantum system describes a precessing qubit chain in the presence of decoherence.
Indeed, in the limit $\theta / \Delta, \theta/\gamma \to \infty$, the qubits are decoupled and individually rotate under the unitary $\mathcal{U}_Z$.
Since the initial state is polarized along the $X$-axis, this leads to $\Tr \prod_j X_j \rho(T) = \cos^L(\theta T)$ while turning on $\Delta, \gamma$ results in decoherence which damps out the oscillations.
Conversely, the state is expected to retain very little coherence in the limit of strong decoherence $\theta/\gamma, \theta/\Delta \to 0$ leading to non-oscillatory decay of $O$ as the state approaches the maximally-mixed state.
The Yang-Lee critical point then describes the threshold $\theta_c$ at which the behavior goes from being underdamped to overdamped.
At this point, the expectation value of $O$ is critically damped.

As a result of our mapping, the same physics can be identified with the exceptional point of the non-Hermitian Hamiltonian $\hat{H}_\mathrm{YL}$ which generates the Yang-Lee transfer matrix.
In particular, the overdamped phase corresponds to the $\mathcal{PT}$-symmetric phase of $\hat{H}_\mathrm{YL}$
where all eigenvalues are real, and consequently the time-dependence of $O$ can only exhibit exponential decay~\eqref{eqn:nhh_projection_dictionary}.
On the other hand, the underdamped phase is the $\mathcal{PT}$ spontaneously-broken phase in which the eigenvalues are generically complex and occur in conjugate pairs.
As a result, the dependence of $O$ exhibits oscillatory behavior with decay set by the real part of the eigenvalues.
Finally, the exceptional point of $\hat{H}_\mathrm{YL}$ is identified with the critically-damped point and the critical point of the classical Yang-Lee theory~\cite{gehlen_yl_as_nhh}. 

We comment that while the underdamped phase can be understood in terms of spontaneous breaking of the $\mathcal{PT}$-symmetry $\mathcal{K} \prod_j \hat{\tau}^z_j$, the symmetry-broken states do not correspond to any physical density matrix.
Indeed, the $\mathcal{PT}$-symmetry of $\hat{H}_\mathrm{YL}$ is nothing more than the $\mathbb{Z}_2$ Hermiticity-preserving symmetry of the Lindbladian~\eqref{eqn:yl_lindbladian_form} projected into the non-Hermitian subspace $\Lambda$.
This is explicitly shown in Appendix~\ref{app:pt_symmetry}.
Thus, physical Hermitian density matrices cannot break the $\mathcal{PT}$-symmetry and instead the symmetry-broken states control the damped-oscillatory decay of observables rather than the steady-state.
Put differently, symmetry-broken states lie within $\Lambda$, have zero trace and are unphysical.

\begin{figure*}[t]
    \centering
    \includegraphics[width=.95\textwidth  ]{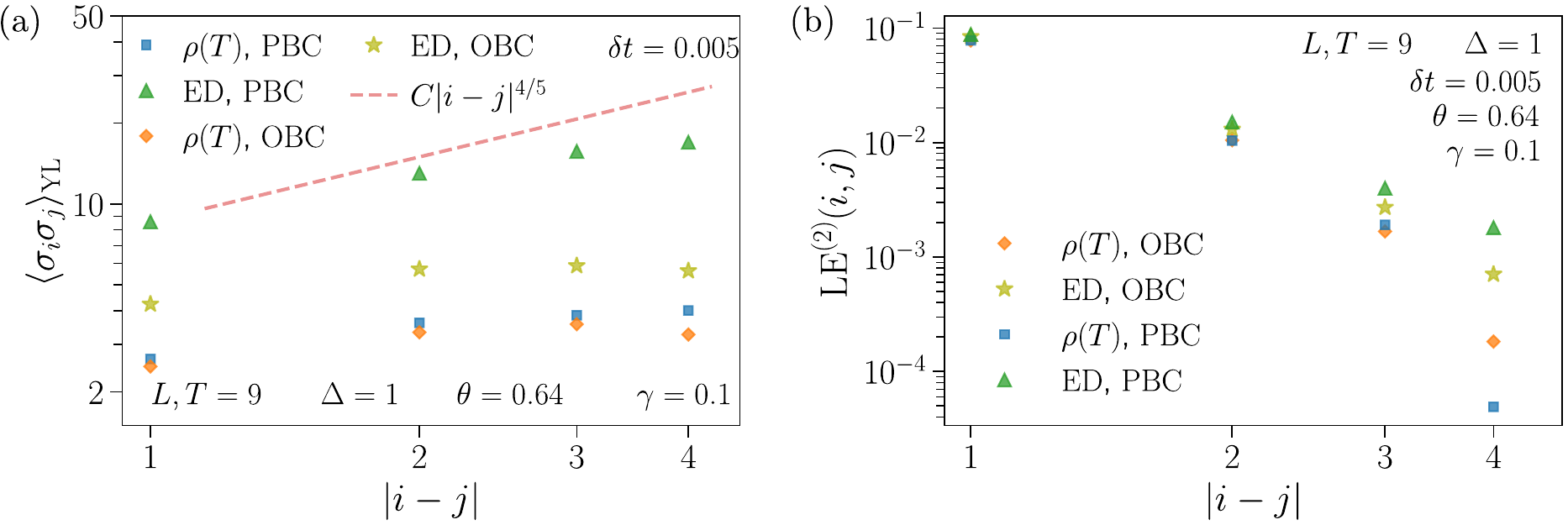}
    \caption{Two-point correlation functions at the critical point $\theta = 0.64, \gamma=0.1, \Delta =1$.
    In (a), 
    we plot $\langle \sigma_i \sigma_j \rangle_\mathrm{YL}$ of the Yang-Lee theory obtained by inserting $V = e^{\ri \pi Z/4}$ at sites $(i, j)$ on an equal-time slice.
    In red, we plot the $\propto |i-j|^{4/5}$ scaling predicted by the Yang-Lee conformal field theory.
    In (b), we plot $\mathrm{LE}^{(2)}(i, j)$, also obtained by inserting $V$ at sites $(i, j)$ at time $t = T/2$ except the bulk dynamics are generated by $\mathcal{L}, \mathcal{L}^\dagger$ for $t < T/2, t>T/2$, respectively.
    In both, $\rho(T)$ (squares) indicates the correlator obtained by directly time-evolving $\rho(T)$ under $\mathcal{L}$ while $\mathrm{ED}$ (triangle, star) indicates the correlator measured in the left/right ground eigenstates of $\hat{H}_\mathrm{YL}$ to which $\rho(T)$ is expected to converge as $T \to \infty$.
    Both open (OBC) and periodic (PBC) boundary conditions are explored.
    }\label{fig:correlators}
\end{figure*}

We verify the physical picture of the Yang-Lee theory as an open quantum system through numerical study of the discrete time dynamics introduced in Sec.~\ref{sec:yl_setup}.
Starting from the initial state $\rho(0) = \ket{+}^L\bra{+}^L$, we observe an over-to-underdamped transition as $\theta$ is increased while fixing $\Delta, \gamma$.
The observable appears to become overdamped at $\theta < \theta_c \approx 0.64$, which is consistent with previous numerical studies of the spectrum of $\hat{H}_\mathrm{YL}$ with these values of $\gamma, \Delta$~\cite{gehlen_yl_as_nhh}.  The results are presented in Figure~\ref{fig:over_to_underdamped}.

Beyond Yang-Lee theory, we expect that exceptional point physics is generally connected to the critically-damped point of observables in the non-Hermitian subspace for Lindbladians which possess an extensive number of weak-symmetries.
Conversely, we show in Sec.~\ref{sec:general_nhh_without_postselection}, that given any non-Hermitian Hamiltonian $\hat{H}_\mathrm{target}$, it is always possible to construct a parent Lindbladian with an extensive number of weak-symmetries which reduces to $\hat{H}_\mathrm{target}$ within a particular weak-symmetry sector.
A precise demonstration of this connection was made in~\cite{yan_ctfim} for a specific Lindbladian whose reduction to a non-Hermitian Hamiltonian was exactly solvable.
\subsection{Spin Correlators and Loschmidt Echos}

The conformal field theory which describes the Yang-Lee critical point hosts a single non-trivial primary operator $\varphi$ with negative scaling dimension $\Delta_\varphi = -2/5$~\cite{Cardy_YL_Theory}.
Conformal invariance fixes the spin-spin correlation function $\langle \sigma(r) \sigma(0) \rangle \propto |r|^{4/5}$ 
which exhibits unbounded growth with separation as a consequence of the complex weights in  Eq.~\eqref{eqn:yl_classical_partition_function}.

We present a concrete protocol for directly measuring spin-spin correlation functions and the associated negative scaling dimensions within our construction.
Recall that we may view the classical Yang-Lee partition function as being defined on the dynamical spacetime history of the open quantum system.
Thus, spin correlation functions in the classical theory are related to the ``sensitivity'' of the expectation value to inserting point defects within the spacetime history at which the dynamics are locally modified.
We make this notion precise below, giving two separate protocols for measuring spin correlators.

In the first protocol, the correlation function is related to the sensitivity of the decay rate of $\Tr O \rho(T)$ to infinitesimal perturbations of the unitary rotation angle.
Indeed, note that the imaginary field strength $h$ is minimally coupled to the spins in $\mathcal{Z}_\mathrm{YL}$ while $-\log \Tr O \rho(T)$ maps to the free energy of the Yang-Lee theory with free boundary conditions.
Consequently, connected correlation functions are obtained by allowing $\theta(x, t)$ to vary in spacetime
\begin{align}
    \langle \sigma_1(\mathbf{r}_1) 
    &\sigma_2(\mathbf{r}_2)
    \cdots \sigma_n(\mathbf{r}_n)\rangle_{\mathrm{conn}}
    \\
    &=
    -
    \left(\ri \delta t\right)^{-n}
    \prod_{k=1}^n
    \partial_{\theta(x_k, t_k)}
    \log \Tr O \rho(T) 
    \Big|_{\theta(x, t) = \theta}
    \,.
    \nonumber
\end{align}
Here, $\mathbf{r} = (i, j)$ is obtained from the spacetime coordinate $(x, t)$ by Wick rotation.
The full correlation function with disconnected piece is also obtained this way by replacing $\log \Tr O \rho(T)$ with $\Tr O \rho(T)$.

On the other hand, the second protocol relates the full correlation function to the
change in $\Tr O \rho(T)$ 
after modifying the dynamics by inserting a finite unitary rotation $V = e^{\ri \pi Z / 4}$.
That is, we define the time-evolved state after the modified dynamics $\tilde{\rho}(T)$
\begin{equation}
    \tilde{\rho}(T) = \hat{T}
    \left(
    \prod_{k=1}^n 
    V_{x_k, t_k}
    \times \Phi_{T/\delta t} \circ \Phi_{T/\delta t}
    \circ \cdots \circ \Phi_1
    \right)
    [\rho(0)]
    \,.
\end{equation}
Here, each $\Phi_t$ are identical and given by Eq.~\eqref{eqn:single_time_step_channel}; the only purpose of the time-ordering symbol $\hat{T}$ is to insert the operators $V_{x_k, t_k}$ at the appropriate times $t_k$.
This protocol is pictured in Fig.~\ref{fig:circit_diagram}b for the two-point spin-spin correlation function.
To identify the corresponding action of $V$ within the non-Hermitian subspace, observe that conjugating by $V$ takes $X \to -Y$ and $Y \to X$ and thus $V$ is identified with $\ri \hat{\tau}^z$.
As a result $(-\ri)^n \Tr O \tilde{\rho}(T)$ is mapped to the Yang-Lee partition function with spin insertions at coordinates corresponding to $(x_k, t_k)$ such that the spin correlation function is given by
\begin{equation}
     \langle \sigma_1(\mathbf{r}_1) \sigma_2(\mathbf{r}_2)\cdots \sigma_n(\mathbf{r}_n)\rangle
    =
    (-\ri)^n
    \frac{\Tr O \tilde{\rho}(T)}{\Tr O \rho(T)}
    \,.
\end{equation}

In Fig.~\ref{fig:correlators}a, we obtain the spin-spin correlation function $\langle \sigma_i \sigma_j \rangle_\mathrm{YL}$ using the second protocol by numerically simulating the discrete time-evolution described in Sec.~\ref{sec:yl_setup}.
The spin-spin correlator is obtained by inserting $V$ at sites $i$ and $j$ at time $t = T/2$.
We observe that $\langle \sigma_i \sigma_j \rangle_\mathrm{YL} > 1$ and appears to grow monotonically with separation $|i-j|$.
Thus, the observed correlator is consistent with the unbounded growth $\propto |i-j|^{4/5}$ predicted by the CFT.
However, we do not observe quantitative agreement with the predicted scaling which we attribute to finite-sized effects.
In the infinite time limit $T\to \infty$, the correlator is given by
\begin{equation}\label{eqn:spin_spin_infinite_time}
    \langle \sigma_i \sigma_j \rangle_\mathrm{YL} \sim 
    \frac{
    \bra{G}_L
    \hat{\tau}^z_i \hat{\tau}^z_j
    \ket{G}_R
    }
    {
    \bra{G}_L
    \ket{G}_R
    }
    \,,
\end{equation}
where $\ket{G}_{L/R}$ denotes the boundary state projected into the left/right groundstate manifold of $\hat{H}_\mathrm{YL}$, spanned by left/right eigenstates which minimize the real part of the associated energies.
In Fig.~\ref{fig:correlators}a, we also determine~\eqref{eqn:spin_spin_infinite_time} after obtaining $\ket{G}_{L/R}$ through exact diagonalization.
We observe significant deviation of the finite-$T$ correlator from the expectation for infinite time, although the qualitative features appear consistent.

While $\langle \sigma_i \sigma_j \rangle_\mathrm{YL}$ is well-motivated from the classical Yang-Lee theory, the following related quantity is natural in the context of open quantum systems
\begin{equation}
\mathrm{LE}^{(n)}(x_1, \dots x_n) :=
    \frac{
    \langle \langle O| 
    e^{T\mathcal{L}^\dagger/2}
    \prod_{k=1}^n
    V_{x_k}
    \,
    e^{T\mathcal{L}/2}
    | \rho(0) \rangle\rangle
    }
    {
    \langle \langle O| 
    e^{T\mathcal{L}^\dagger/2}
    e^{T\mathcal{L}/2}
    | \rho(0) \rangle\rangle
    }
    \,.
\end{equation}
This quantity measures the expectation value of $O$ after imperfect time-reversal due to the defects $V_{x_k}$ and resembles a Loschmidt echo~\cite{loschmidt_echo_review,loschmidt_dqpt} but which differs from the standard definition in the following two regards.
First, while the standard Loschmidt echo typically characterizes the fidelity of a state following an imperfect time-reversal, $\mathrm{LE}^{(n)}$ is defined with respect to expectation values of a particular operator $O \in \Lambda$.
Second, our definition normalizes by the decaying expectation value of $O$ in the absence of defect insertions and should be thought of as a generalization of the usual Loschmidt echo in which the unperturbed dynamics are given by a known Lindbladian.

In Fig.~\ref{fig:correlators}b, we numerically determine $\mathrm{LE}^{(2)}$ which appears to be bounded above by $1$ and a decaying function of separation $|i-j|$.
Indeed, in the limit $T \to \infty$, we have
\begin{equation}\label{eqn:loschmidt_echo_infinite_time}
    \mathrm{LE}^{(2)}(i, j) \sim 
    \frac{
    \bra{G}_R
    \hat{\tau}^z_i \hat{\tau}^z_j
    \ket{G}_R
    }
    {
    \bra{G}_R
    \ket{G}_R
    }
    \,,
\end{equation}
which is manifestly bounded.
Again, we compare our numerics with Eq.~\eqref{eqn:loschmidt_echo_infinite_time} determined by exact diagonalization.
Our result appears to be largely independent on the boundary conditions while the finite-time $\mathrm{LE}^{(2)}$ appears close to the infinite-time value.

In the effective continuum field theory~\cite{Fisher_YL_Theory}, the Loschmidt echo correlator can be understood in terms of correlation functions measured along a parity defect across which the order parameter field $\varphi \to - \varphi$ \textit{without} complex conjugation.
This takes the form of a problem in boundary conformal field theory.
Since $\mathrm{LE}^{(2)}$ is bounded, on general grounds we expect powerlaw correlators with positive scaling dimension.\footnote{The complex weights in principle allow for another scenario in which the boundary spontaneously breaks the symmetry.  However, our numerics suggest that the spin-spin correlation functions are powerlaw decaying.}
It would be interesting to develop a theoretical understanding of the expected scaling of these objects.

Finally, previous proposals have also suggested measuring the effective central charge, which appears as the universal constant term $c_\mathrm{eff} = 2/5$ of the free energy scaling in finite-sized systems~\cite{probing_central_charge}.
We comment that this quantity can also be measured in our setup, since the decay rate $-\log \Tr O \rho(T)$ is directly related to the free energy of the corresponding classical Yang-Lee theory.

\section{General non-Hermitian Hamiltonians without Post-Selection}\label{sec:general_nhh_without_postselection}
In this section, we present a general and exact method to realize non-Hermitian physics without the use of post-selection.
Our construction generalizes the strategy used to realize the Yang-Lee partition function in Sec.~\ref{sec:non_hermitian_primer}.
Specifically, we will show that given any non-Hermitian Hamiltonian $\hat{H}_\mathrm{target}$, there exists a Lindbladian $\mathcal{L}$ and associated non-Hermitian subspace $\Lambda$ such that $\mathcal{L}$ projected into $\Lambda$ is equivalent to $\hat{H}_\mathrm{target}$.

In particular, the time-evolution generated by $\hat{H}_\mathrm{target}$ is then identified as the effective description for the relaxation of expectation values of observables $O \in \Lambda$ in a mixed-state evolving under the open quantum dynamics described by $\mathcal{L}$.
In particular, we have
\begin{equation}\label{eqn:general_dictionary}
    \Tr\left( O e^{T \mathcal{L}}[\rho(0)]\right) = 
    \bra{\phi_f} e^{-T \hat{H}_\mathrm{target}} \ket{\phi_i} \,,
\end{equation}
where $\ket{\phi_i}\cong P_\Lambda | \rho(0)\rangle\rangle$ and $\ket{\phi_f} \cong P_\Lambda |O \rangle\rangle$ are the initial and final states determined by the initial density matrix $\rho(0)$ and observable $O$ projected into the non-Hermitian subspace $\Lambda$.
We expect that if $\hat{H}_\mathrm{target}$ possess a $\mathcal{PT}$-symmetry which is spontaneously broken at an exceptional point as a set of parameters $\vec{g}$ are tuned, the dynamics of the corresponding observable~\eqref{eqn:general_dictionary} will undergo an overdamped-to-underdamped transition.

Our construction makes two assumptions about the form of $\hat{H}_\mathrm{target}$:
\begin{itemize}
    \item $\hat{H}_\mathrm{target}$ acts on Hilbert space with a tensor product structure, i.e., it can be viewed as a lattice Hamiltonian but with potentially all-to-all couplings.
    \item
    We assume that the local Hilbert space on each lattice site is finite dimensional.
    For simplicity, we assume that the dimension $d$ on all sites is the same, but this can be relaxed.
\end{itemize}
Subject to these conditions, the desired Lindbladian $\mathcal{L}$ can always be found, but it acts on a potentially larger open system.
In particular, for each lattice site, we  introduce $n = \left\lceil \log_2 d \right\rceil$ independent qubits and an additional auxillary qudit with a $4$-dimensional Hilbert space.
We will use the $n$ qubits to emulate a single $d$-dimensional qudit degree of freedom while the $4$-dimensional qudit is required for the most general form of $\hat{H}_\mathrm{target}$.\footnote{
For simplicity, we work with qubits.
In principle, only a single $4$-dimensional auxillary qudit is required, but introducing a qudit for each lattice site allows us to keep the Lindbladian local if $\hat{H}_\mathrm{target}$ is also local.
  Other constructions using qudit or bosonic degrees of freedom instead of qubits are also possible.
}
\subsection{Non-Hermiticity from Jump Operators}\label{sec:general_constraints}
In this section, we overview the strategy for constructing the desired Lindbladian which we are able to accomplish by only engineering the jump operator terms.

This approach differs from other approaches, such as when non-Hermiticity is obtained from Lindbladian evolution using post-selection or $\hat{H}_\mathrm{YL}$ obtained by projecting into $\Lambda$ in Sec.~\ref{sec:non_hermitian_primer} and~\ref{sec:yang_lee_construction}.  
In both cases, the unitary part of the Lindbladian generates real-time evolution while the jump operator terms generate imaginary-time evolution.
In the latter, if one considers the Lindbladian obtained by 
dropping either set of terms, the resulting projected dynamics results in (real or imaginary) time-evolution governed by a Hermitian Hamiltonian.

The motivation for relying on jump operator engineering is that, while the unitary part of $\mathcal{L}$ can be useful for generating non-Hermiticity, its action in $\Lambda$ is essentially restricted to on-site terms.
This is due to the fact that the Lie bracket is a derivation.
As a result, given a multi-site operator $W = \prod_i W_i$, its action on $\sigma \in \Lambda$ is given by
\begin{equation}
    \left[
        W, \sigma
    \right]
    = \sum_i 
    \left( \prod_{j < i} W_j\left[\mathcal{O}_i, \sigma
    \right]
    \prod_{k>i} W_k
    \right)
    \,.
\end{equation}
In other words, even if we consider multi-site unitaries, the action projected into $\Lambda$ is a sum of single-site terms, limiting its applicability for engineering general Hermiticity-breaking terms.

Conversely, the jump operators act on $\sigma \in \Lambda$ as
\begin{equation}
    \sigma \to \sum_\alpha L_\alpha^\dagger \sigma L_\alpha - \frac{1}{2}\{L^\dagger_\alpha L_\alpha, \sigma\} \,.
\end{equation}
In what follows, we will take the $L_\alpha$ to be proportional to a unitary.
Then, the jump operators act by conjugation whose action projected into $\Lambda$ remains quite general.
Unitary of $L_\alpha$ guarantees that the jump operator-only Lindbladian preserves Hermiticity, while a constant term must be added to preserve the trace, which we drop for simplicity.

Next, we construct the non-Hermitian subspace $\Lambda$.
The subspace will be spanned by operators of the form
\begin{equation}\label{eqn:general_construction_nhh_subspace}
    \prod_i \left(
    C_i \prod_\mu O_{i, \mu} \,,
    \right)
\end{equation}
with $O_{i, \mu} = X_{i, \mu}, Y_{i, \mu}$ a two-dimensional Pauli matrix and where $i = 1, \dots , N$ denote the site index and $\alpha = 1, \dots, n$ the flavor index.
$C_i$ and $S_i$ are the four-dimensional clock and shift matrices acting on the ancillary qudit associated with site $i$.
The matrices satisfy $C_j^4 = S_j^4 = I_j$ and $C_j S_j = \ri S_j C_j$ for all $j$.

The main idea behind our construction can now be stated as follows:  we choose jump operators $L_\alpha$ proportional to Clifford unitaries whose action by conjugation either permute the basis states~\eqref{eqn:general_construction_nhh_subspace} of $\Lambda$, or stabilize them up to a phase~\cite{gottesman_stabilizer,clifford_calderbank}.
Furthermore, for our construction based on qubits, the ancillary qudit is introduced because the relation 
$S^\dagger C S = \ri C$ implies that we may always append $S$ to a jump operator $L_\alpha$ such that its action projected into $\Lambda$ then gains a factor of $\ri$ realizing non-Hermiticity.

Finally, as in Sec.~\ref{sec:non_hermitian_primer}, the observable $O$ must be chosen to lie within $\Lambda$ while the initial density matrix must be chosen to have a finite overlap with $\Lambda$.
In this case, one can choose $O = \prod_i C_i \prod_\mu X_{i, \mu}$.
The initial state can be chosen to be $\rho(0) = \ket{\ri}^N \ket{+}^{nN} \bra{\ri}^N \bra{+}^{nN} $, i.e., the state which stabilizes $-\ri C_j$ and $X_{j, \mu}$ for all $j, \mu$.
\subsection{Explicit  Construction}\label{sec:generic_construction}
The key observation we use is that we are able to
construct the desired Lindbladian in steps by first 
 constructing jump operators which reproduce particular matrix elements of $\hat{H}_\mathrm{target}$.
 Then, the full Lindbladian which reproduces the entirety of $\hat{H}_\mathrm{target}$ is obtained by collecting the set of all jump operators which were required for each matrix element.
 This follows from linearity of the Lindbladian and $\hat{H}_\mathrm{target}$.

Let $\hat{H}_\mathrm{target}$ decompose as
\begin{equation}
    \hat{H}_\mathrm{target} = \sum_{\sigma} \hat{H}_\sigma \,,
\end{equation}
with each $\hat{H}_\sigma$ acting on $K$-nearest neighbors.
This assumption is completely general and can be relaxed, since we do not use the locality property of $\hat{H}_\sigma$ in the following.

From the above, it suffices to construct jump operators which reduce to $\hat{H}_\sigma$ after projecting into $\Lambda$.
The most general form of $\hat{H}_\sigma$ subject to our constraints takes the form
\begin{equation}\label{eqn:general_NHH}
    \hat{H}_\sigma = \sum_{\vec{e}, \vec{e}^{\, '}}
    H_{\vec{e}, 
    \vec{e}^{\,'}}
    |
    \vec{e}
    \rangle
    \langle
    \vec{e}^{\, '}
    |
     \,.
\end{equation}
Here $\vec{e} = (e_1, e_2, \dots, e_K) \in \Z_d^K$ enumerates a basis for the Hilbert space on which $\hat{H}_\sigma$ acts non-trivially.

We will emulate each $d$-dimensional qudit degree of freedom with $n = \left\lceil \log_2 d \right\rceil$ qubits.
To this end, it will be useful to define the following non-Hermitian Hamiltonian, which is equivalent to $\hat{H}_{\sigma}$
\begin{equation}\label{eqn:general_NHH_bit_string}
    \hat{H}_{\sigma, \mathrm{qubit}} = \sum_{\vec{b}, \vec{b}^{\, '}}
    H_{\vec{b}, \vec{b}^{\, '}}
    |
    \vec{b}
    \rangle
    \langle
    \vec{b}^{\, '}
    |
    \,,
\end{equation}
where $\vec{b} \in \mathbb{Z}_2^{n K}$ is a fixed bit-string representation of $\vec{e}$.
In other words, any qudit Hamiltonian is reducible to a qubit Hamiltonian up to a constant multiplicative overhead in the number of qubits and locality of the interactions.

The basis for the non-Hermitian subspace~\eqref{eqn:general_construction_nhh_subspace} are mapped onto the states $| \vec{b} \rangle$ by associating the spin-up state of each qubit with $X$ and the spin-down state with $Y$.
Formally,
\begin{equation}
|\vec{b}\rangle \to
    \prod_j
    \left(
    C_j 
    \prod_\mu
    X_{j, \mu}
    (\ri Z_{j, \mu})^{b_{j, \mu}}
    \right)
    \,.
\end{equation}
While the $\vec{b}$ appearing here is a basis for the subspace on which $\hat{H}_\sigma$ acts non-trivially, it is clear that it may be consistently extended to a complete basis for the entire system.

In the following, we construct jump operators which reproduce particular matrix elements $ H_{\vec{b}, \vec{b}^{\, '}}
    |
    \vec{b}
    \rangle
    \langle
    \vec{b}^{\, '}
    |$ of $\hat{H}_{\sigma, \mathrm{qubit}}$.
    Again, the construction for the entirety of $\hat{H}_\mathrm{qubit}$ and $\hat{H}_\mathrm{target}$ follows by linearity.
In particular, we show that $\mathcal{O}(D)$ jump operators are required for each matrix element leading to $\mathcal{O}(D^3)$ total jump operators where $D = 2^{nK}$ is the Hilbert space dimension.
Since a generic Lindbladian has at most $\mathcal{O}(D^2)$ independent jump operators, this construction does not necessarily provide the most efficient or physically natural representation of $\mathcal{L}$.

\subsubsection{Diagonal Matrix Elements}
First, we consider diagonal matrix elements of $\hat{H}_{\sigma,\mathrm{qubit}}$
\begin{equation}
H_{\vec{b}, \vec{b}}
|
\vec{b}
\rangle
\langle
\vec{b}
|
    \,,
\end{equation}
which are proportional to the projector onto the bit string $\vec{b}$.  
Without loss of generality, we assume 
that the prefactor $H_{\vec{b}, \vec{b}}$ is either purely real or purely imaginary.
The general case follows by linearity.

The projector onto a specific bit string $\vec{b}$ can be written as a sum over Pauli matrices $\sigma^z$ acting on the $|\vec{b}\rangle$ states weighted by appropriate signs
\begin{equation}\label{eqn:bit_string_projector}
|
\vec{b}
\rangle
\langle
\vec{b}
|
    = 
    \frac{1}{2^{nK}}
    \sum_{\vec{c} \in \Z^{nK}_2}
    \left(
    -1
    \right)^{\vec{b} \cdot \vec{c}}
    \times
    \prod_{\mu, j} \left(
    \sigma^z_{j, \mu}
    \right)^{c_{j, \mu}}
    \,.
\end{equation}

Next, observe that $X_{j, \mu}$ commutes with $X_{j, \mu}$ and anti-commutes with $Y_{j, \mu}$ and thus the action $\sigma \to X_{j, \mu} \sigma X_{j, \mu}$ is equivalent to $\sigma^z_{j, \mu}$ for $\sigma \in \Lambda$.
Thus, in order to realize projectors on $\Lambda$, we define the $2^{nK}$ jump operators labeled by $\vec{c} \in \Z_2^{nK}$
\begin{equation}
L_{\vec{c}} =
    \frac{
    \sqrt{|H_{\vec{b}, \vec{b}}|}
    }
    {
    2^{nK / 2}
    }
    S_s^\beta
    S_s^{2 \vec{b} \cdot \vec{c}}
    \prod_{j, \mu} X_{j, \mu}^{c_{j, \mu}}
    \,.
\end{equation}
Here, $S_s$ is the shift operator defined on one of the ancillary qudits.
If spatial locality of $\mathcal{L}$ is desired, the site index $s$ is chosen to be one of the sites $j$ on which $\hat{H}_{\sigma, \mathrm{qubit}}$ has support.
Next, conjugation by $L_{\vec{c}}$ realizes the action of the corresponding term in Eq.~\eqref{eqn:bit_string_projector} on $\Lambda$.
The term $\prod_{j, \mu} X_{j, \mu}^{c_{j, \mu}}$ 
maps to $\prod_{j, \mu} (\sigma^z_{j, \mu})^{c_{j, \mu}}$ while 
$S_s^{2 \vec{b} \cdot \vec{c}}$ generates the sign $(-1)^{\vec{b} \cdot \vec{c}}$, using the fact that $S_s^2 C_s S_s^2 = - C_s$.
Finally, since $S_s^\dagger C_s S_s = \ri C_s$, we choose $\beta = 0, 1, 2, 3$ in the case that $H_{\vec{b}, \vec{b}}$ is positive, positive imaginary, negative or negative imaginary, respectively.

\subsubsection{Off-Diagonal Matrix Elements}
Next, we construct jump operators for a general off-diagonal matrix element, which can be written as
\begin{equation}
H_{\vec{b}, \vec{b}^{\, '}} 
|
\vec{b}
\rangle
\langle
\vec{b}^{\, '}
|
=
H_{\vec{b}, \vec{b}^{\, '}} 
\times
\left(
|
\vec{b}
\rangle
\langle
\vec{b}^{\, '}
|
+
|
\vec{b}^{\, '}
\rangle
\langle
\vec{b}
|
\right)
\times
|
\vec{b}^{\, '}
\rangle
\langle
\vec{b}^{\, '}
|
    \,,
\end{equation}
which we have written as a projector into $| \vec{b}^{\, '}\rangle$ followed by bit flips $\sigma^x$ which map the bit strings $\vec{b}^{\, '}$ and $\vec{b}$ into each other.
Again, we assume $H_{\vec{b}, \vec{b}^{\, '}} $ is either purely real or purely imaginary.
The series of bit flips can be expressed as
\begin{equation}
|
\vec{b}
\rangle
\langle
\vec{b}^{\, '}
|
+
|
\vec{b}^{\, '}
\rangle
\langle
\vec{b}
|
= \prod_{j, \mu}
(\sigma^x_{j, \mu})^{b_{j, \mu} \oplus b^{\, '}_{j, \mu}}
\,,
\end{equation}
where $\oplus$ denotes addition mod $2$.

Next, conjugation by $\ri e^{-\ri \frac{\pi}{2\sqrt{2} }(Y+ X)} = \frac{1}{\sqrt{2}} \left( X + Y\right)$ 
acts as $\sigma^x$ on $\Lambda$ since it sends $X \leftrightarrow Y$.
Thus, our $2^{nK}$ jump operators parameterized by $\vec{c} \in \Z_{2}^{nK}$ take the form
\begin{align}
L_{\vec{c}} =
    \frac{
    \sqrt{|H_{\vec{b}, \vec{b}}|}
    }
    {
    2^{nK / 2}
    }
    S_s^{2 \vec{c} \cdot \vec{b}^{\, '} + \beta}
    \prod_{j, \mu} 
    \left(
    \frac{X_{j, \mu} + Y_{j, \mu}}{\sqrt{2}}
    \right)^{b_{j, \mu}\oplus b_{j, \mu}^{\, '}}
    &
    \nonumber
    \\
    \times\prod_{j, \mu} X_{j, \mu}^{c_{j, \mu}}&
    \,.
\end{align}
Again, with $\beta = 0, 1, 2, 3$ depending on the reality and sign of $H_{\vec{b}, \vec{b}^{\, '}}$.

\section{Discussion}\label{sec:conclusion}
In this work, we studied the Lindbladian dynamics of an open quantum system which possess an extensive number of local weak-symmetries and its reduction to the pure state time-evolution generated by a non-Hermitian Hamiltonian.
In this setting, non-Hermiticity arises as the effective description, made precise through an exact mapping, governing the relaxation of expectation values of a certain class of observables which live in a \textit{non-Hermitian subspace}.
Our result connects the phenomena of over-to-underdamped transitions~\cite{yan_ctfim} 
in the time-evolution of such observables under continuous open quantum dynamics
 with the study of dynamical quantum phase transitions in quantum quenches~\cite{Heyl_DQPT_review,Heyl_DQPT_TFIM,Jurcevic_DQPT_observation,Flaschner_DQPT_cold_atom,Budich_DQPT_topological,Sharma_DQPT_topological} associated with exceptional points in non-Hermitian quantum mechanics~\cite{Kato_Perturbation_Theory_Lin_Operators,Heiss_Physics_Exceptional_Points,ashida_non_hermitian_review,Heiss_Level_Crossing,Heiss_Phases_Of_Wavefunctions,Moiseyev_NH_QM}.

Our main result is the explicit construction of a Lindbladian whose action on certain observables in the non-Hermitian subspace is exactly equivalent to that of quantum states evolving under the non-Hermitian Hamiltonian generating the infinitesimal transfer matrix of Yang-Lee theory.
In particular, the expectation value of the observables is equal to the Yang-Lee partition function defined on the spacetime history of the dynamics.
Modifying the dynamics locally in spacetime allows direct measurement of spin correlation functions, while tuning the dynamics to criticality gives an experimental probe of the Yang-Lee conformal critical point in any spatial dimension.

Further, we present a discrete-time formulation of our protocol amenable for implementation on digital quantum devices, equivalent up to Suzuki-Trotter errors.
Our construction involves only nearest-neighbor gates and dissipative channels engineered through controlled interactions with ancillae serving as ``bath'' degrees of freedom.
Consequently, it is tailored to quantum simulator platforms that allow for highly-tunable interactions and single-site control such as Rydberg tweezer arrays~\cite{saffman_rydberg,gross_qgm,browaeys_many_body}.
In particular, in addition to single-site phase gates, the only two-qubit gates required are of the CNOT and CZ type, which have been experimentally demonstrated on this platform.
We must also perform measurements of the non-local, string-like observable $\prod_i X_i$.
This can be reconstructed by the statistics of a site-resolved readout.

Additionally, we are able to generalize this construction.
For a wide class of non-Hermitian Hamiltonians, we explicitly demonstrate the existence of a parent Lindbladian which possess an extensive set of weak-symmetries.
The Lindbladian reduces to the non-Hermitian Hamiltonian in a particular weak-symmetry sector, which may be fixed by considering the expectation value of a particular observable in the time-evolved mixed-state.

Our work points to a few directions for future research.
For instance, our construction for general non-Hermitian Hamiltonians motivates the search for 
lattice realizations of non-unitary theories which are less well-understood and can be realized on quantum simulator devices in a similar setup.
One example are the multi-critical Yang-Lee theories which has been the subject of recent theoretical  study~\cite{gehlen_tricritical,lencs_multicritical,katsevich_gl_multicritical,lencs_gl_multicritical}.
Similarly, our construction could be applied to experimental studies of Yang-Lee criticality in higher-dimensions complementing recent numerical advances~\cite{yl_fuzzy_sphere_various_d,fan_fuzzy_sphere}.

It is also interesting to ask if our construction could be further generalized.
In particular, we rely on an extensive number of local ``weak-symmetries'' which implicitly assumes the tensor product structure of  Hilbert space.
We are interested in exploring if this assumption can be relaxed, enabling the simulation of a wider range of non-unitary models~\cite{yl_anyons} whose unitary counterparts have been proposed to be realizable in Rydberg atom arrays~\cite{fendley_rydberg}.
Next, our work improves upon previous results in its lack of post-selection, which would otherwise incur an exponential overhead in sampling complexity.
However, we comment that measuring $\prod_i X_i$, which is proportional to the partition function, would still require an exponential number of experimental runs to resolve in practice.
It would be interesting to find a realization of Yang-Lee theory which does not rely on measuring the partition function and avoid this issue.
Finally, the projection into the non-Hermitian subspace formally requires the measurement of non-local observables.
We wonder if non-Hermitian physics can be realized without post-selection in a fully local setting. 

\begin{acknowledgements}
We thank Jacob Hauser, Thomas Kiely, Andreas W.W. Ludwig, Rushikesh A. Patil, Shinsei Ryu, Masahito Ueda, Sagar Vijay and David M. Weld for helpful discussions.
    S.W.Y. acknowledges the support of the National Science Foundation under Grant No. DMR–2441671.
    S.W.Y. was supported by a grant from the W. M. Keck Foundation.
    This research was done using services provided by the OSG Consortium~\cite{osg_1,osg_2,osg_3,osg_4}, which is supported by the National Science Foundation awards No. 2030508 and No. 2323298.
\end{acknowledgements}

\bibliography{biblio}

@article{Minganti_NHH_no_go,
   title={Quantum exceptional points of non-Hermitian Hamiltonians and Liouvillians: The effects of quantum jumps},
   volume={100},
   ISSN={2469-9934},
   url={http://dx.doi.org/10.1103/PhysRevA.100.062131},
   DOI={10.1103/physreva.100.062131},
   number={6},
   journal={Physical Review A},
   publisher={American Physical Society (APS)},
   author={Minganti, Fabrizio and Miranowicz, Adam and Chhajlany, Ravindra W. and Nori, Franco},
   year={2019},
   month=dec }

@article{Matsumoto_YL_exp,
  title = {Embedding the Yang-Lee quantum criticality in open quantum systems},
  author = {Matsumoto, Norifumi and Nakagawa, Masaya and Ueda, Masahito},
  journal = {Phys. Rev. Res.},
  volume = {4},
  issue = {3},
  pages = {033250},
  numpages = {15},
  year = {2022},
  month = {Sep},
  publisher = {American Physical Society},
  doi = {10.1103/PhysRevResearch.4.033250},
  url = {https://link.aps.org/doi/10.1103/PhysRevResearch.4.033250}
}

@article{Gao_YL_exp,
  title = {Experimental Observation of the Yang-Lee Quantum Criticality in Open Quantum Systems},
  author = {Gao, Huixia and Wang, Kunkun and Xiao, Lei and Nakagawa, Masaya and Matsumoto, Norifumi and Qu, Dengke and Lin, Haiqing and Ueda, Masahito and Xue, Peng},
  journal = {Phys. Rev. Lett.},
  volume = {132},
  issue = {17},
  pages = {176601},
  numpages = {6},
  year = {2024},
  month = {Apr},
  publisher = {American Physical Society},
  doi = {10.1103/PhysRevLett.132.176601},
  url = {https://link.aps.org/doi/10.1103/PhysRevLett.132.176601}
}

@misc{Abo_YL_exp,
      title={Liouvillian Exceptional Points of Non-Hermitian Systems via Quantum Process Tomography}, 
      author={Shilan Abo and Patrycja Tulewicz and Karol Bartkiewicz and Şahin K. Özdemir and Adam Miranowicz},
      year={2024},
      eprint={2401.14993},
      archivePrefix={arXiv},
      primaryClass={quant-ph},
      url={https://arxiv.org/abs/2401.14993}, 
}

@article{Cardy_YL_Theory,
  title = {Conformal Invariance and the Yang-Lee Edge Singularity in Two Dimensions},
  author = {Cardy, John L.},
  journal = {Phys. Rev. Lett.},
  volume = {54},
  issue = {13},
  pages = {1354--1356},
  numpages = {0},
  year = {1985},
  month = {Apr},
  publisher = {American Physical Society},
  doi = {10.1103/PhysRevLett.54.1354},
  url = {https://link.aps.org/doi/10.1103/PhysRevLett.54.1354}
}

@article{Fisher_YL_Theory,
  title = {Yang-Lee Edge Singularity and ${\ensuremath{\phi}}^{3}$ Field Theory},
  author = {Fisher, Michael E.},
  journal = {Phys. Rev. Lett.},
  volume = {40},
  issue = {25},
  pages = {1610--1613},
  numpages = {0},
  year = {1978},
  month = {Jun},
  publisher = {American Physical Society},
  doi = {10.1103/PhysRevLett.40.1610},
  url = {https://link.aps.org/doi/10.1103/PhysRevLett.40.1610}
}

@article{Yang_Lee_YL_Theory_I,
  title = {Statistical Theory of Equations of State and Phase Transitions. I. Theory of Condensation},
  author = {Yang, C. N. and Lee, T. D.},
  journal = {Phys. Rev.},
  volume = {87},
  issue = {3},
  pages = {404--409},
  numpages = {0},
  year = {1952},
  month = {Aug},
  publisher = {American Physical Society},
  doi = {10.1103/PhysRev.87.404},
  url = {https://link.aps.org/doi/10.1103/PhysRev.87.404}
}

@article{Yang_Lee_YL_Theory_II,
  title = {Statistical Theory of Equations of State and Phase Transitions. II. Lattice Gas and Ising Model},
  author = {Lee, T. D. and Yang, C. N.},
  journal = {Phys. Rev.},
  volume = {87},
  issue = {3},
  pages = {410--419},
  numpages = {0},
  year = {1952},
  month = {Aug},
  publisher = {American Physical Society},
  doi = {10.1103/PhysRev.87.410},
  url = {https://link.aps.org/doi/10.1103/PhysRev.87.410}
}

@article{Wei_BB_Central_Spin,
  title = {Lee-Yang Zeros and Critical Times in Decoherence of a Probe Spin Coupled to a Bath},
  author = {Wei, Bo-Bo and Liu, Ren-Bao},
  journal = {Phys. Rev. Lett.},
  volume = {109},
  issue = {18},
  pages = {185701},
  numpages = {5},
  year = {2012},
  month = {Oct},
  publisher = {American Physical Society},
  doi = {10.1103/PhysRevLett.109.185701},
  url = {https://link.aps.org/doi/10.1103/PhysRevLett.109.185701}
}

@article{experimental_peng,
   title={Experimental Observation of Lee-Yang Zeros},
   volume={114},
   ISSN={1079-7114},
   url={http://dx.doi.org/10.1103/PhysRevLett.114.010601},
   DOI={10.1103/physrevlett.114.010601},
   number={1},
   journal={Physical Review Letters},
   publisher={American Physical Society (APS)},
   author={Peng, Xinhua and Zhou, Hui and Wei, Bo-Bo and Cui, Jiangyu and Du, Jiangfeng and Liu, Ren-Bao},
   year={2015},
   month=Jan }

@article{Francis_Central_Spin,
author = {Akhil Francis  and Daiwei Zhu  and Cinthia Huerta Alderete  and Sonika Johri  and Xiao Xiao  and James K. Freericks  and Christopher Monroe  and Norbert M. Linke  and Alexander F. Kemper },
title = {Many-body thermodynamics on quantum computers via partition function zeros},
journal = {Science Advances},
volume = {7},
number = {34},
pages = {eabf2447},
year = {2021},
doi = {10.1126/sciadv.abf2447},
URL = {https://www.science.org/doi/abs/10.1126/sciadv.abf2447},
eprint = {https://www.science.org/doi/pdf/10.1126/sciadv.abf2447}}

@book{Fradkin_field_theory,
    author = "Fradkin, Eduardo",
    title = "{Quantum Field Theory: An Integrated Approach}",
    isbn = "978-0-691-14908-0",
    publisher = "Princeton University Press",
    month = "3",
    year = "2021"
}

@inproceedings{osg_1,
  title  = {The open science grid},
  author = {
    Pordes, Ruth
    and Petravick, Don
    and Kramer, Bill
    and Olson, Doug
    and Livny, Miron
    and Roy, Alain
    and Avery, Paul
    and Blackburn, Kent
    and Wenaus, Torre
    and W{\"u}rthwein, Frank
    and Foster, Ian
    and Gardner, Rob
    and Wilde, Mike
    and Blatecky, Alan
    and McGee, John
    and Quick, Rob
  },
  doi       = {10.1088/1742-6596/78/1/012057},
  booktitle = {J. Phys. Conf. Ser.},
  volume    = {78},
  series    = {78},
  pages     = {012057},
  year      = {2007},
}

@inproceedings{osg_2,
  title        = {The pilot way to grid resources using glideinWMS},
  author       = {
    Sfiligoi, Igor
    and Bradley, Daniel C
    and Holzman, Burt
    and Mhashilkar, Parag
    and Padhi, Sanjay
    and Wurthwein, Frank
  },
  doi          = {10.1109/CSIE.2009.950},
  booktitle    = {2009 WRI World Congress on Computer Science and Information Engineering},
  volume       = {2},
  series       = {2},
  pages        = {428--432},
  year         = {2009},
}

@misc{osg_3,
  doi = {10.21231/906P-4D78},
  url = {https://osg-htc.org/services/open_science_pool.html},
  author = {{OSG}},
  title = {OSPool},
  publisher = {OSG},
  year = {2006}
}

@misc{osg_4,
  doi = {10.21231/0KVZ-VE57},
  url = {https://osdf.osg-htc.org/},
  author = {{OSG}},
  title = {Open Science Data Federation},
  publisher = {OSG},
  year = {2015}
}

@article{gehlen_yl_as_nhh,
doi = {10.1088/0305-4470/24/22/021},
url = {https://doi.org/10.1088/0305-4470/24/22/021},
year = {1991},
month = {nov},
publisher = {},
volume = {24},
number = {22},
pages = {5371},
author = {G von Gehlen},
title = {Critical and off-critical conformal analysis of the Ising quantum chain in an imaginary field},
journal = {Journal of Physics A: Mathematical and General}
}

@article{fonseca_yl_scaling,
    author = "Fonseca, P. and Zamolodchikov, A.",
    title = "{Ising field theory in a magnetic field: Analytic properties of the free energy}",
    eprint = "hep-th/0112167",
    archivePrefix = "arXiv",
    reportNumber = "RUNHETC-2001-37",
    month = "12",
    year = "2001"
}

@article{mangazeev_yl_critical_location,
   title={Corner transfer matrix approach to the Yang-Lee singularity in the two-dimensional Ising model in a magnetic field},
   volume={108},
   ISSN={2470-0053},
   url={http://dx.doi.org/10.1103/PhysRevE.108.064136},
   DOI={10.1103/physreve.108.064136},
   number={6},
   journal={Physical Review E},
   publisher={American Physical Society (APS)},
   author={Mangazeev, Vladimir V. and Hagan, Bryte and Bazhanov, Vladimir V.},
   year={2023},
   month=Dec }

@article{kurtze_yl_spherical,
  author  = {Kurtze, Douglas A. and Fisher, Michael E.},
  title   = {The Yang-Lee edge singularity in spherical models},
  journal = {Journal of Statistical Physics},
  year    = {1978},
  volume  = {19},
  number  = {3},
  pages   = {205--218},
  doi     = {10.1007/BF01011723},
  issn    = {1572-9613}
}

@article{fisher_yl_1d,
    author = {Fisher, Michael E.},
    title = {Yang-Lee Edge Behavior in One-Dimensional Systems},
    journal = {Progress of Theoretical Physics Supplement},
    volume = {69},
    pages = {14-29},
    year = {1980},
    month = {03},
    issn = {0375-9687},
    doi = {10.1143/PTP.69.14},
    url = {https://doi.org/10.1143/PTP.69.14},
    eprint = {https://academic.oup.com/ptps/article-pdf/doi/10.1143/PTP.69.14/5348942/69-14.pdf},
}

@article{Buca_Prosen_Weak_Symmetry,
   title={A note on symmetry reductions of the Lindblad equation: transport in constrained open spin chains},
   volume={14},
   ISSN={1367-2630},
   url={http://dx.doi.org/10.1088/1367-2630/14/7/073007},
   DOI={10.1088/1367-2630/14/7/073007},
   number={7},
   journal={New Journal of Physics},
   publisher={IOP Publishing},
   author={Buča, Berislav and Prosen, Tomaž},
   year={2012},
   month=jul, pages={073007}
}

@article{Albert_Jiang_Weak_Symmetry,
   title={Symmetries and conserved quantities in Lindblad master equations},
   volume={89},
   ISSN={1094-1622},
   url={http://dx.doi.org/10.1103/PhysRevA.89.022118},
   DOI={10.1103/physreva.89.022118},
   number={2},
   journal={Physical Review A},
   publisher={American Physical Society (APS)},
   author={Albert, Victor V. and Jiang, Liang},
   year={2014},
   month=feb
}

@misc{yan_ctfim,
      title={Dissipative Dynamical Phase Transition as a Complex Ising Model}, 
      author={Stephen W. Yan and Diego Barberena and Matthew P. A. Fisher and Sagar Vijay},
      year={2024},
      eprint={2412.09591},
      archivePrefix={arXiv},
      primaryClass={quant-ph},
      url={https://arxiv.org/abs/2412.09591}, 
}

@article{loschmidt_echo_review,
   title={Loschmidt echo},
   volume={7},
   ISSN={1941-6016},
   url={http://dx.doi.org/10.4249/scholarpedia.11687},
   DOI={10.4249/scholarpedia.11687},
   number={8},
   journal={Scholarpedia},
   publisher={Scholarpedia},
   author={Wisniacki, Arseni},
   year={2012},
   pages={11687} }

@article{loschmidt_dqpt,
  title = {Theory of the Loschmidt echo and dynamical quantum phase transitions in disordered Fermi systems},
  author = {Vanhala, Tuomas I. and Ojanen, Teemu},
  journal = {Phys. Rev. Res.},
  volume = {5},
  issue = {3},
  pages = {033178},
  numpages = {7},
  year = {2023},
  month = {Sep},
  publisher = {American Physical Society},
  doi = {10.1103/PhysRevResearch.5.033178},
  url = {https://link.aps.org/doi/10.1103/PhysRevResearch.5.033178}
}

@article{probing_central_charge,
   title={Probing Conformal Invariant of Non-unitary Two-Dimensional Systems by Central Spin Decoherence},
   volume={8},
   ISSN={2045-2322},
   url={http://dx.doi.org/10.1038/s41598-018-21360-7},
   DOI={10.1038/s41598-018-21360-7},
   number={1},
   journal={Scientific Reports},
   publisher={Springer Science and Business Media LLC},
   author={Wei, Bo-Bo},
   year={2018},
   month=Feb }

@article{wei_central_decoherence_proposal,
   title={Probing Yang–Lee edge singularity by central spin decoherence},
   volume={19},
   ISSN={1367-2630},
   url={http://dx.doi.org/10.1088/1367-2630/aa77d6},
   DOI={10.1088/1367-2630/aa77d6},
   number={8},
   journal={New Journal of Physics},
   publisher={IOP Publishing},
   author={Wei, Bo-Bo},
   year={2017},
   month=Aug, pages={083009}
}

@article{Heyl_DQPT_review,
   title={Dynamical quantum phase transitions: a review},
   volume={81},
   ISSN={1361-6633},
   url={http://dx.doi.org/10.1088/1361-6633/aaaf9a},
   DOI={10.1088/1361-6633/aaaf9a},
   number={5},
   journal={Reports on Progress in Physics},
   publisher={IOP Publishing},
   author={Heyl, Markus},
   year={2018},
   month=apr, pages={054001} }

@article{Heyl_DQPT_TFIM,
  title = {Dynamical Quantum Phase Transitions in the Transverse-Field Ising Model},
  author = {Heyl, M. and Polkovnikov, A. and Kehrein, S.},
  journal = {Phys. Rev. Lett.},
  volume = {110},
  issue = {13},
  pages = {135704},
  numpages = {5},
  year = {2013},
  month = {Mar},
  publisher = {American Physical Society},
  doi = {10.1103/PhysRevLett.110.135704},
  url = {https://link.aps.org/doi/10.1103/PhysRevLett.110.135704}
}

@article{Budich_DQPT_topological,
   title={Dynamical topological order parameters far from equilibrium},
   volume={93},
   ISSN={2469-9969},
   url={http://dx.doi.org/10.1103/PhysRevB.93.085416},
   DOI={10.1103/physrevb.93.085416},
   number={8},
   journal={Physical Review B},
   publisher={American Physical Society (APS)},
   author={Budich, Jan Carl and Heyl, Markus},
   year={2016},
   month=feb }

@article{Sharma_DQPT_topological,
   title={Slow quenches in a quantum Ising chain: Dynamical phase transitions and topology},
   volume={93},
   ISSN={2469-9969},
   url={http://dx.doi.org/10.1103/PhysRevB.93.144306},
   DOI={10.1103/physrevb.93.144306},
   number={14},
   journal={Physical Review B},
   publisher={American Physical Society (APS)},
   author={Sharma, Shraddha and Divakaran, Uma and Polkovnikov, Anatoli and Dutta, Amit},
   year={2016},
   month=apr }

@article{Flaschner_DQPT_cold_atom,
   title={Observation of dynamical vortices after quenches in a system with topology},
   volume={14},
   ISSN={1745-2481},
   url={http://dx.doi.org/10.1038/s41567-017-0013-8},
   DOI={10.1038/s41567-017-0013-8},
   number={3},
   journal={Nature Physics},
   publisher={Springer Science and Business Media LLC},
   author={Fläschner, N. and Vogel, D. and Tarnowski, M. and Rem, B. S. and Lühmann, D.-S. and Heyl, M. and Budich, J. C. and Mathey, L. and Sengstock, K. and Weitenberg, C.},
   year={2017},
   month=dec, pages={265–268} }

@article{Jurcevic_DQPT_observation,
   title={Direct Observation of Dynamical Quantum Phase Transitions in an Interacting Many-Body System},
   volume={119},
   ISSN={1079-7114},
   url={http://dx.doi.org/10.1103/PhysRevLett.119.080501},
   DOI={10.1103/physrevlett.119.080501},
   number={8},
   journal={Physical Review Letters},
   publisher={American Physical Society (APS)},
   author={Jurcevic, P. and Shen, H. and Hauke, P. and Maier, C. and Brydges, T. and Hempel, C. and Lanyon, B. P. and Heyl, M. and Blatt, R. and Roos, C. F.},
   year={2017},
   month=aug }

@book{Kato_Perturbation_Theory_Lin_Operators,
  alias = {Kato 66},
  author = {Kato, Tosio},
  bibdate = {Fri Nov 24 15:18:30 1995},
  bibsource = {ftp://ftp.math.utah.edu/pub/bibnet/authors/m/matched-field-proc.bib},
  key = {eigenvalues, hilbert spaces},
  lccn = {QA320 .K33},
  pages = {xix + 592},
  sthbib = {M3 Kat 81 60},
publisher={Springer Berlin Heidelberg},
  title = {Perturbation Theory for Linear Operators},
  year = 1966
}

@article{Heiss_Physics_Exceptional_Points,
    author = "Heiss, W. D.",
    title = "{The physics of exceptional points}",
    eprint = "1210.7536",
    archivePrefix = "arXiv",
    primaryClass = "quant-ph",
    doi = "10.1088/1751-8113/45/44/444016",
    journal = "J. Phys. A",
    volume = "45",
    pages = "444016",
    year = "2012"
}

@article{Heiss_Level_Crossing,
doi = {10.1088/0305-4470/23/7/022},
url = {https://dx.doi.org/10.1088/0305-4470/23/7/022},
year = {1990},
month = {apr},
publisher = {},
volume = {23},
number = {7},
pages = {1167},
author = {W D Heiss and  A L Sannino},
title = {Avoided level crossing and exceptional points},
journal = {Journal of Physics A: Mathematical and General}
}

@article{Heiss_Phases_Of_Wavefunctions,
    author = "Heiss, W. D.",
    title = "{Phases of wave functions and level repulsion}",
    eprint = "quant-ph/9901023",
    archivePrefix = "arXiv",
    doi = "10.1007/s100530050339",
    journal = "Eur. Phys. J. D",
    volume = "7",
    pages = "1",
    year = "1999"
}

@book{Moiseyev_NH_QM,
  title={Non-Hermitian Quantum Mechanics},
  author={Moiseyev, N.},
  isbn={9780511992124},
  url={https://books.google.com/books?id=p9LYzgEACAAJ},
  year={2011},
  publisher={Cambridge University Press}
}

@article{ashida_non_hermitian_review,
   title={Non-Hermitian physics},
   volume={69},
   ISSN={1460-6976},
   url={http://dx.doi.org/10.1080/00018732.2021.1876991},
   DOI={10.1080/00018732.2021.1876991},
   number={3},
   journal={Advances in Physics},
   publisher={Informa UK Limited},
   author={Ashida, Yuto and Gong, Zongping and Ueda, Masahito},
   year={2020},
   month=July, pages={249–435} }

@article{trotter,
 ISSN = {00029939, 10886826},
 URL = {http://www.jstor.org/stable/2033649},
 author = {H. F. Trotter},
 journal = {Proceedings of the American Mathematical Society},
 number = {4},
 pages = {545--551},
 publisher = {American Mathematical Society},
 title = {On the Product of Semi-Groups of Operators},
 urldate = {2026-08-05},
 volume = {10},
 year = {1959}
}

@article{suzuki,
  author  = {Suzuki, Masuo},
  title   = {Generalized {T}rotter's formula and systematic approximants of exponential operators and inner derivations with applications to many-body problems},
  journal = {Communications in Mathematical Physics},
  year    = {1976},
  volume  = {51},
  number  = {2},
  pages   = {183--190},
  doi     = {10.1007/BF01609348},
  issn    = {1432-0916}
}

@article{gehlen_tricritical,
author = {GEHLEN, G. VON},
title = {NON-HERMITIAN TRICRITICALITY IN THE BLUME-CAPEL MODEL WITH IMAGINARY FIELD},
journal = {International Journal of Modern Physics B},
volume = {08},
number = {25n26},
pages = {3507-3529},
year = {1994},
doi = {10.1142/S0217979294001494},
URL = { 
        https://doi.org/10.1142/S0217979294001494
},
eprint = { 
        https://doi.org/10.1142/S0217979294001494
}
}

@article{lencs_multicritical,
   title={Multicriticality in Yang-Lee edge singularity},
   volume={2023},
   ISSN={1029-8479},
   url={http://dx.doi.org/10.1007/JHEP02(2023)046},
   DOI={10.1007/jhep02(2023)046},
   number={2},
   journal={Journal of High Energy Physics},
   publisher={Springer Science and Business Media LLC},
   author={Lencsés, Máté and Miscioscia, Alessio and Mussardo, Giuseppe and Takács, Gábor},
   year={2023},
   month=Feb }

@article{katsevich_gl_multicritical,
    author = "Katsevich, Andrei and Klebanov, Igor R. and Sun, Zimo and Tarnopolsky, Grigory",
    title = "{Towards a Quintic Ginzburg-Landau Description of the (2,7) Minimal Model}",
    eprint = "2510.19085",
    archivePrefix = "arXiv",
    primaryClass = "hep-th",
    doi = "10.1103/9zz4-pvp7",
    journal = "Phys. Rev. Lett.",
    volume = "136",
    number = "11",
    pages = "111602",
    year = "2026"
}

@article{lencs_gl_multicritical,
    author = "Lencs{\'e}s, M{\'a}t{\'e} and Miscioscia, Alessio and Mussardo, Giuseppe and Tak{\'a}cs, G{\'a}bor",
    title = "{Ginzburg-Landau description for multicritical Yang-Lee models}",
    eprint = "2404.06100",
    archivePrefix = "arXiv",
    primaryClass = "cond-mat.stat-mech",
    reportNumber = "DESY-24-032",
    doi = "10.1007/JHEP08(2024)224",
    journal = "JHEP",
    volume = "08",
    pages = "224",
    year = "2024"
}

@article{yl_fuzzy_sphere_various_d,
  title = {Yang-Lee Quantum Criticality in Various Dimensions},
  author = {Arguello Cruz, Erick and Klebanov, Igor R. and Tarnopolsky, Grigory and Xin, Yuan},
  journal = {Phys. Rev. X},
  volume = {16},
  issue = {1},
  pages = {011022},
  numpages = {27},
  year = {2026},
  month = {Feb},
  publisher = {American Physical Society},
  doi = {10.1103/w4qg-2xwn},
  url = {https://link.aps.org/doi/10.1103/w4qg-2xwn}
}

@article{fan_fuzzy_sphere,
    author = "Fan, Ruihua and Dong, Junkai and Vishwanath, Ashvin",
    title = "{Simulating the non-unitary Yang-Lee conformal field theory on the fuzzy sphere}",
    eprint = "2505.06342",
    archivePrefix = "arXiv",
    primaryClass = "cond-mat.str-el",
    month = "5",
    year = "2025"
}

@article{yl_anyons,
   title={Microscopic models of interacting Yang–Lee anyons},
   volume={13},
   ISSN={1367-2630},
   url={http://dx.doi.org/10.1088/1367-2630/13/4/045006},
   DOI={10.1088/1367-2630/13/4/045006},
   number={4},
   journal={New Journal of Physics},
   publisher={IOP Publishing},
   author={Ardonne, E and Gukelberger, J and Ludwig, A W W and Trebst, S and Troyer, M},
   year={2011},
   month=Apr, pages={045006} }

@article{fendley_rydberg,
  title = {Microscopic characterization of Ising conformal field theory in Rydberg chains},
  author = {Slagle, Kevin and Aasen, David and Pichler, Hannes and Mong, Roger S. K. and Fendley, Paul and Chen, Xie and Endres, Manuel and Alicea, Jason},
  journal = {Phys. Rev. B},
  volume = {104},
  issue = {23},
  pages = {235109},
  numpages = {20},
  year = {2021},
  month = {Dec},
  publisher = {American Physical Society},
  doi = {10.1103/PhysRevB.104.235109},
  url = {https://link.aps.org/doi/10.1103/PhysRevB.104.235109}
}

@article{lindblad_dyn_semigroup,
author = {G. Lindblad},
title = {{On the generators of quantum dynamical semigroups}},
volume = {48},
journal = {Communications in Mathematical Physics},
number = {2},
publisher = {Springer},
pages = {119 -- 130},
year = {1976},
}

@article{gorini_dyn_semigroup,
    author = "Gorini, Vittorio and Kossakowski, Andrzej and Sudarshan, E. C. G.",
    title = "{Completely Positive Dynamical Semigroups of N Level Systems}",
    reportNumber = "CPT-244-TEXAS, ORO-3992-200",
    doi = "10.1063/1.522979",
    journal = "J. Math. Phys.",
    volume = "17",
    pages = "821",
    year = "1976"
}

@article{gottesman_stabilizer,
    author = "Gottesman, Daniel",
    title = "{Stabilizer codes and quantum error correction}",
    eprint = "quant-ph/9705052",
    archivePrefix = "arXiv",
    month = "5",
    year = "1997"
}

@article{clifford_calderbank,
    author = "Calderbank, A. R. and Rains, E. M. and Shor, P. W. and Sloane, N. J. A.",
    title = "{Quantum error correction via codes over GF(4)}",
    eprint = "quant-ph/9608006",
    archivePrefix = "arXiv",
    doi = "10.1109/18.681315",
    journal = "IEEE Trans. Info. Theor.",
    volume = "44",
    number = "4",
    pages = "1369--1387",
    year = "1998"
}

@article{Bender_PT_Sym,
  title = {Real Spectra in Non-Hermitian Hamiltonians Having $\mathcal{P}\mathcal{T}$ Symmetry},
  author = {Bender, Carl M. and Boettcher, Stefan},
  journal = {Phys. Rev. Lett.},
  volume = {80},
  issue = {24},
  pages = {5243--5246},
  numpages = {0},
  year = {1998},
  month = {Jun},
  publisher = {American Physical Society},
  doi = {10.1103/PhysRevLett.80.5243},
  url = {https://link.aps.org/doi/10.1103/PhysRevLett.80.5243}
}

@article{Mostafazadeh_PT_Sym,
   title={Pseudo-Hermitian Representation of Quantum Mechanics},
   volume={07},
   ISSN={1793-6977},
   url={http://dx.doi.org/10.1142/S0219887810004816},
   DOI={10.1142/s0219887810004816},
   number={07},
   journal={International Journal of Geometric Methods in Modern Physics},
   publisher={World Scientific Pub Co Pte Lt},
   author={Mostafazadeh, Ali},
   year={2010},
   month=nov, pages={1191–1306} }

@article{friedan_minimal,
  author  = {Friedan, D. and Qiu, Z. and Shenker, S.},
  title   = {Conformal Invariance, Unitarity, and Critical Exponents in Two Dimensions},
  journal = {Physical Review Letters},
  year    = {1984},
  volume  = {52},
  number  = {18},
  pages   = {1575--1578},
  doi     = {10.1103/PhysRevLett.52.1575}
}

@article{belavin_minimal,
  author  = {Belavin, A. A. and Polyakov, A. M. and Zamolodchikov, A. B.},
  title   = {Infinite conformal symmetry in two-dimensional quantum field theory},
  journal = {Nuclear Physics B},
  year    = {1984},
  volume  = {241},
  number  = {2},
  pages   = {333--380},
  doi     = {10.1016/0550-3213(84)90052-X}
}

@article{deger_fluctuations,
  title = {Lee-Yang theory, high cumulants, and large-deviation statistics of the magnetization in the Ising model},
  author = {Deger, Aydin and Brange, Fredrik and Flindt, Christian},
  journal = {Phys. Rev. B},
  volume = {102},
  issue = {17},
  pages = {174418},
  numpages = {12},
  year = {2020},
  month = {Nov},
  publisher = {American Physical Society},
  doi = {10.1103/PhysRevB.102.174418},
  url = {https://link.aps.org/doi/10.1103/PhysRevB.102.174418}
}

@article{binek_anal_cont,
  title = {Density of Zeros on the Lee-Yang Circle Obtained from Magnetization Data of a Two-Dimensional Ising Ferromagnet},
  author = {Binek, Ch.},
  journal = {Phys. Rev. Lett.},
  volume = {81},
  issue = {25},
  pages = {5644--5647},
  numpages = {0},
  year = {1998},
  month = {Dec},
  publisher = {American Physical Society},
  doi = {10.1103/PhysRevLett.81.5644},
  url = {https://link.aps.org/doi/10.1103/PhysRevLett.81.5644}
}

@article{essler2020integrability,
  title={Integrability of one-dimensional Lindbladians from operator-space fragmentation},
  url = {https://link.aps.org/doi/10.1103/PhysRevE.102.062210},
  author={Essler, Fabian HL and Piroli, Lorenzo},
  journal={Physical Review E},
  volume={102},
  number={6},
  pages={062210},
  year={2020},
  publisher={APS}
}

@article{PhysRevResearch.5.043239,
  title = {Hilbert space fragmentation in open quantum systems},
  author = {Li, Yahui and Sala, Pablo and Pollmann, Frank},
  journal = {Phys. Rev. Res.},
  volume = {5},
  issue = {4},
  pages = {043239},
  numpages = {17},
  year = {2023},
  month = {Dec},
  publisher = {American Physical Society},
  doi = {10.1103/PhysRevResearch.5.043239},
  url = {https://link.aps.org/doi/10.1103/PhysRevResearch.5.043239}
}

@article{PhysRevB.109.054311,
  title = {Hidden quasilocal charges and Gibbs ensemble in a Lindblad system},
  author = {de Leeuw, Marius and Paletta, Chiara and Pozsgay, Bal\'azs and Vernier, Eric},
  journal = {Phys. Rev. B},
  volume = {109},
  issue = {5},
  pages = {054311},
  numpages = {13},
  year = {2024},
  month = {Feb},
  publisher = {American Physical Society},
  doi = {10.1103/PhysRevB.109.054311},
  url = {https://link.aps.org/doi/10.1103/PhysRevB.109.054311}
}

@article{de_Groot_2022,
   title={Symmetry Protected Topological Order in Open Quantum Systems},
   volume={6},
   ISSN={2521-327X},
   url={http://dx.doi.org/10.22331/q-2022-11-10-856},
   DOI={10.22331/q-2022-11-10-856},
   journal={Quantum},
   publisher={Verein zur Forderung des Open Access Publizierens in den Quantenwissenschaften},
   author={de Groot, Caroline and Turzillo, Alex and Schuch, Norbert},
   year={2022},
   month=Nov, pages={856} }

@article{Ma_2023,
   title={Average Symmetry-Protected Topological Phases},
   volume={13},
   ISSN={2160-3308},
   url={http://dx.doi.org/10.1103/PhysRevX.13.031016},
   DOI={10.1103/physrevx.13.031016},
   number={3},
   journal={Physical Review X},
   publisher={American Physical Society (APS)},
   author={Ma, Ruochen and Wang, Chong},
   year={2023},
   month=Aug }

@article{Diehl_2008,
   title={Quantum states and phases in driven open quantum systems with cold atoms},
   volume={4},
   ISSN={1745-2481},
   url={http://dx.doi.org/10.1038/nphys1073},
   DOI={10.1038/nphys1073},
   number={11},
   journal={Nature Physics},
   publisher={Springer Science and Business Media LLC},
   author={Diehl, S. and Micheli, A. and Kantian, A. and Kraus, B. and Büchler, H. P. and Zoller, P.},
   year={2008},
   month=Sept, pages={878–883} }

@article{Ma_2025,
   title={Topological Phases with Average Symmetries: The Decohered, the Disordered, and the Intrinsic},
   volume={15},
   ISSN={2160-3308},
   url={http://dx.doi.org/10.1103/PhysRevX.15.021062},
   DOI={10.1103/physrevx.15.021062},
   number={2},
   journal={Physical Review X},
   publisher={American Physical Society (APS)},
   author={Ma, Ruochen and Zhang, Jian-Hao and Bi, Zhen and Cheng, Meng and Wang, Chong},
   year={2025},
   month=May }

@article{Verstraete_2008,
    author = "Verstraete, Frank and Wolf, Michael M. and Cirac, J. Ignacio",
    title = "{Quantum computation, quantum state engineering, and quantum phase transitions driven by dissipation}",
    eprint = "0803.1447",
    archivePrefix = "arXiv",
    primaryClass = "quant-ph",
    journal = "Nature Phys.",
    volume = "5",
    pages = "633--636",
    year = "2009"
}

@article{Fan_2024,
   title={Diagnostics of Mixed-State Topological Order and Breakdown of Quantum Memory},
   volume={5},
   ISSN={2691-3399},
   url={http://dx.doi.org/10.1103/PRXQuantum.5.020343},
   DOI={10.1103/prxquantum.5.020343},
   number={2},
   journal={PRX Quantum},
   publisher={American Physical Society (APS)},
   author={Fan, Ruihua and Bao, Yimu and Altman, Ehud and Vishwanath, Ashvin},
   year={2024},
   month=May }

@article{Lee_2023,
   title={Quantum Criticality Under Decoherence or Weak Measurement},
   volume={4},
   ISSN={2691-3399},
   url={http://dx.doi.org/10.1103/PRXQuantum.4.030317},
   DOI={10.1103/prxquantum.4.030317},
   number={3},
   journal={PRX Quantum},
   publisher={American Physical Society (APS)},
   author={Lee, Jong Yeon and Jian, Chao-Ming and Xu, Cenke},
   year={2023},
   month=Aug }

@article{Lee_2022,
    author = "Lee, Jong Yeon and Ji, Wenjie and Bi, Zhen and Fisher, Matthew P. A.",
    title = "{Decoding Measurement-Prepared Quantum Phases and Transitions: from Ising model to gauge theory, and beyond}",
    eprint = "2208.11699",
    archivePrefix = "arXiv",
    primaryClass = "cond-mat.str-el",
    month = "8",
    year = "2022"
}

@article{Lu_2023,
   title={Mixed-State Long-Range Order and Criticality from Measurement and Feedback},
   volume={4},
   ISSN={2691-3399},
   url={http://dx.doi.org/10.1103/PRXQuantum.4.030318},
   DOI={10.1103/prxquantum.4.030318},
   number={3},
   journal={PRX Quantum},
   publisher={American Physical Society (APS)},
   author={Lu, Tsung-Cheng and Zhang, Zhehao and Vijay, Sagar and Hsieh, Timothy H.},
   year={2023},
   month=Aug }

@article{Garratt_2023,
   title={Measurements Conspire Nonlocally to Restructure Critical Quantum States},
   volume={13},
   ISSN={2160-3308},
   url={http://dx.doi.org/10.1103/PhysRevX.13.021026},
   DOI={10.1103/physrevx.13.021026},
   number={2},
   journal={Physical Review X},
   publisher={American Physical Society (APS)},
   author={Garratt, Samuel J. and Weinstein, Zack and Altman, Ehud},
   year={2023},
   month=May }

@article{Zhu_2023,
   title={Nishimori’s Cat: Stable Long-Range Entanglement from Finite-Depth Unitaries and Weak Measurements},
   volume={131},
   ISSN={1079-7114},
   url={http://dx.doi.org/10.1103/PhysRevLett.131.200201},
   DOI={10.1103/physrevlett.131.200201},
   number={20},
   journal={Physical Review Letters},
   publisher={American Physical Society (APS)},
   author={Zhu, Guo-Yi and Tantivasadakarn, Nathanan and Vishwanath, Ashvin and Trebst, Simon and Verresen, Ruben},
   year={2023},
   month=Nov }

@article{Wang_2025,
   title={Intrinsic Mixed-State Topological Order},
   volume={6},
   ISSN={2691-3399},
   url={http://dx.doi.org/10.1103/PRXQuantum.6.010314},
   DOI={10.1103/prxquantum.6.010314},
   number={1},
   journal={PRX Quantum},
   publisher={American Physical Society (APS)},
   author={Wang, Zijian and Wu, Zhengzhi and Wang, Zhong},
   year={2025},
   month=Jan }

@article{P_tz_2025,
   title={Flow to Nishimori Universality in Weakly Monitored Quantum Circuits with Qubit Loss},
   volume={6},
   ISSN={2691-3399},
   url={http://dx.doi.org/10.1103/ygfz-crvp},
   DOI={10.1103/ygfz-crvp},
   number={4},
   journal={PRX Quantum},
   publisher={American Physical Society (APS)},
   author={Pütz, Malte and Vasseur, Romain and Ludwig, Andreas W.W. and Trebst, Simon and Zhu, Guo-Yi},
   year={2025},
   month=Dec }

@article{Jian_2023,
    author = "Jian, Chao-Ming and Shapourian, Hassan and Bauer, Bela and Ludwig, Andreas W. W.",
    title = "{Measurement-induced entanglement transitions in quantum circuits of non-interacting fermions: Born-rule versus forced measurements}",
    eprint = "2302.09094",
    archivePrefix = "arXiv",
    primaryClass = "cond-mat.stat-mech",
    month = "2",
    year = "2023"
}

@article{Yan_2026,
    author = "Yan, Stephen W. and Bao, Yimu and Vijay, Sagar",
    title = "{Non-linear Sigma Model for the Surface Code with Coherent Errors}",
    eprint = "2603.25665",
    archivePrefix = "arXiv",
    primaryClass = "cond-mat.stat-mech",
    month = "3",
    year = "2026"
}

@article{Wang_2025_self_dual,
    author = "Wang, Qingyuan and Vasseur, Romain and Trebst, Simon and Ludwig, Andreas W. W. and Zhu, Guo-Yi",
    title = "{Decoherence-induced self-dual criticality in topological states of matter}",
    eprint = "2502.14034",
    archivePrefix = "arXiv",
    primaryClass = "quant-ph",
    month = "2",
    year = "2025"
}

@article{Skinner_2019,
   title={Measurement-Induced Phase Transitions in the Dynamics of Entanglement},
   volume={9},
   ISSN={2160-3308},
   url={http://dx.doi.org/10.1103/PhysRevX.9.031009},
   DOI={10.1103/physrevx.9.031009},
   number={3},
   journal={Physical Review X},
   publisher={American Physical Society (APS)},
   author={Skinner, Brian and Ruhman, Jonathan and Nahum, Adam},
   year={2019},
   month=July }

@article{Li_2018,
   title={Quantum Zeno effect and the many-body entanglement transition},
   volume={98},
   ISSN={2469-9969},
   url={http://dx.doi.org/10.1103/PhysRevB.98.205136},
   DOI={10.1103/physrevb.98.205136},
   number={20},
   journal={Physical Review B},
   publisher={American Physical Society (APS)},
   author={Li, Yaodong and Chen, Xiao and Fisher, Matthew P. A.},
   year={2018},
   month=Nov }

@article{Chan_2019,
   title={Unitary-projective entanglement dynamics},
   volume={99},
   ISSN={2469-9969},
   url={http://dx.doi.org/10.1103/PhysRevB.99.224307},
   DOI={10.1103/physrevb.99.224307},
   number={22},
   journal={Physical Review B},
   publisher={American Physical Society (APS)},
   author={Chan, Amos and Nandkishore, Rahul M. and Pretko, Michael and Smith, Graeme},
   year={2019},
   month=June }

@article{Patil_2024,
    author = "Patil, Rushikesh A. and Ludwig, Andreas W. W.",
    title = "{Highly complex novel critical behavior from the intrinsic randomness of quantum mechanical measurements on critical ground states -- a controlled renormalization group analysis}",
    eprint = "2409.02107",
    archivePrefix = "arXiv",
    primaryClass = "cond-mat.stat-mech",
    month = "9",
    year = "2024"
}

@article{krishnan_2019,
    author = "Krishnan, Abijith and Schmitt, Markus and Moessner, Roderich and Heyl, Markus",
    title = "{Measuring complex partition function zeroes of Ising models in quantum simulators}",
    eprint = "1902.07155",
    archivePrefix = "arXiv",
    primaryClass = "quant-ph",
    doi = "10.1103/PhysRevA.100.022125",
    journal = "Phys. Rev. A",
    volume = "100",
    pages = "022125",
    year = "2019"
}

@article{browaeys_many_body,
   title={Many-body physics with individually controlled Rydberg atoms},
   volume={16},
   ISSN={1745-2481},
   url={http://dx.doi.org/10.1038/s41567-019-0733-z},
   DOI={10.1038/s41567-019-0733-z},
   number={2},
   journal={Nature Physics},
   publisher={Springer Science and Business Media LLC},
   author={Browaeys, Antoine and Lahaye, Thierry},
   year={2020},
   month=Jan, pages={132–142} 
}

@article{saffman_rydberg,
   title={Quantum information with Rydberg atoms},
   volume={82},
   ISSN={1539-0756},
   url={http://dx.doi.org/10.1103/RevModPhys.82.2313},
   DOI={10.1103/revmodphys.82.2313},
   number={3},
   journal={Reviews of Modern Physics},
   publisher={American Physical Society (APS)},
   author={Saffman, M. and Walker, T. G. and Mølmer, K.},
   year={2010},
   month=Aug, pages={2313–2363} 
}

@article{gross_qgm,
  author  = {Gross, Christian and Bloch, Immanuel},
  title   = {Quantum simulations with ultracold atoms in optical lattices},
  journal = {Science},
  year    = {2017},
  volume  = {357},
  number  = {6355},
  pages   = {995--1001},
  doi     = {10.1126/science.aal3837}
}

@article{harrington_engineered_dissipation,
   title={Engineered dissipation for quantum information science},
   volume={4},
   ISSN={2522-5820},
   url={http://dx.doi.org/10.1038/s42254-022-00494-8},
   DOI={10.1038/s42254-022-00494-8},
   number={10},
   journal={Nature Reviews Physics},
   publisher={Springer Science and Business Media LLC},
   author={Harrington, Patrick M. and Mueller, Erich J. and Murch, Kater W.},
   year={2022},
   month=Aug, pages={660–671} 
}

\onecolumngrid
\newpage 

\appendix

\setcounter{equation}{0}
\setcounter{figure}{0}
\renewcommand{\thetable}{S\arabic{table}}
\renewcommand{\theequation}{S\thesection.\arabic{equation}}
\renewcommand{\thefigure}{S\arabic{figure}}
\setcounter{secnumdepth}{2}

\begin{center}
{\Large Supplementary Material \\
\vspace{0.2cm}
}
\end{center}

\section{Effective Spacelike Ising Interaction in Yang-Lee Theory}\label{sec:app_circ_realization}
Given the desired $2$-site jump operator term in the Lindbladian~\eqref{eqn:2site_lindbladian_operator}, we may guess that the corresponding $2$-site Kraus operator is proportional to
\begin{equation}
    K_{\mu_{i, j}} 
    \sim
    \exp \mu \sqrt{\gamma \delta t} W_i^\dagger W_j
    = 
    \exp \mu \frac{\sqrt{\gamma \delta t} }{2} 
    \left(
        I_i I_j + Z_i Z_j  + \ri I_i Z_j - \ri Z_i I_j
    \right)
    \,,
\end{equation}
where $W_j = \exp(\ri \pi Z_j/4)$.
The action of the quantum channel is obtained by summing over Kraus operators $\mu = \pm 1$
\begin{equation}
\begin{split}
    \sum_\mu 
    &= K_\mu \otimes K_\mu^*
    \\
    &=
    \exp
    \sqrt{\gamma \delta t}
    W_i^\dagger W_j
    \otimes
    \exp
    \sqrt{\gamma \delta t}
    W_i W_j^\dagger
    +
    \exp
    -\sqrt{\gamma \delta t}
    W_i^\dagger W_j
    \otimes
    \exp
    -\sqrt{\gamma \delta t}
    W_i W_j^\dagger
    \\
    &=
    2 
    \cosh 
    \left[
    \sqrt{\gamma \delta t}
    (
    W_i^\dagger W_j
    \otimes I 
    +
    I \otimes
    W_i W_j^\dagger
    )
    \right]
\,.
\end{split}
\end{equation}
This result tells us that in order to fix the normalization, we should divide out by
\begin{equation}
    \sum_\mu K^\dagger_\mu  K_\mu = 2 \cosh 
    \sqrt{\gamma \delta t}
    W_i^\dagger W_j
    +
    W_i W_j^\dagger
    = 
    2 \cosh \sqrt{\gamma \delta t} (I + Z_i Z_j)
    \,.
\end{equation}
Explicitly, 
\begin{equation}\label{eqn:app_normalized_kraus_operator}
    K_{\mu_{i, j}} =  \frac{
    \exp
    \mu
    \sqrt{\gamma \delta t} ( I + Z_i Z_j + \ri Z_j - \ri Z_i) / 2
    }
    {
    \sqrt{
    2 \cosh \sqrt{\gamma \delta t} (I + Z_i Z_j)
    }
    }
    \,.
\end{equation}
The normalized quantum channel
\begin{equation}
\begin{split}
    \mathcal{K}_{ZZ}
    &=
    \sum_\mu K_\mu \otimes K_\mu^* 
    \\
    &\approx
    \frac{1}{2} \sum_\mu
    \left[I + \mu \sqrt{\gamma \delta t} W_i^\dagger W_j - \frac{\gamma \delta t}{2} + \mathcal{O}\left((\gamma \delta t)^{3/2}\right) \right]
    \otimes
    \left[I + \mu \sqrt{\gamma \delta t} W_i W_j^\dagger - \frac{\gamma \delta t}{2} + \mathcal{O}\left((\gamma \delta t)^{3/2}\right) \right]
    \\
    &= \frac{1}{2}
    \sum_{\mu}
    \left[
    I + \mu \sqrt{\gamma \delta t}\left(W_i^\dagger W_j \otimes I + I \otimes W_i W_j^\dagger\right)
    - \gamma \delta t + \gamma \delta t W_i^\dagger W_j \otimes 
    W_i W_j^\dagger
    +
    \mathcal{O}\left((\gamma \delta t)^{3/2}\right)
    \right]
    \\
    & \approx
    I + 
 \gamma \delta t
 \left(
W_i^\dagger W_j \otimes 
    W_i W_j^\dagger
        -  I
        \right)
    \approx e^{\gamma \delta t (W_i^\dagger W_j \otimes 
    W_i W_j^\dagger - I)}
    \,,
\end{split}
\end{equation}
gives us the desired form~\eqref{eqn:2site_lindbladian_operator} up to order $\delta t$ and dropping constant normalization factors.

\begin{figure*}[t]
    \centering
    \includegraphics[width=.96 \textwidth  ]{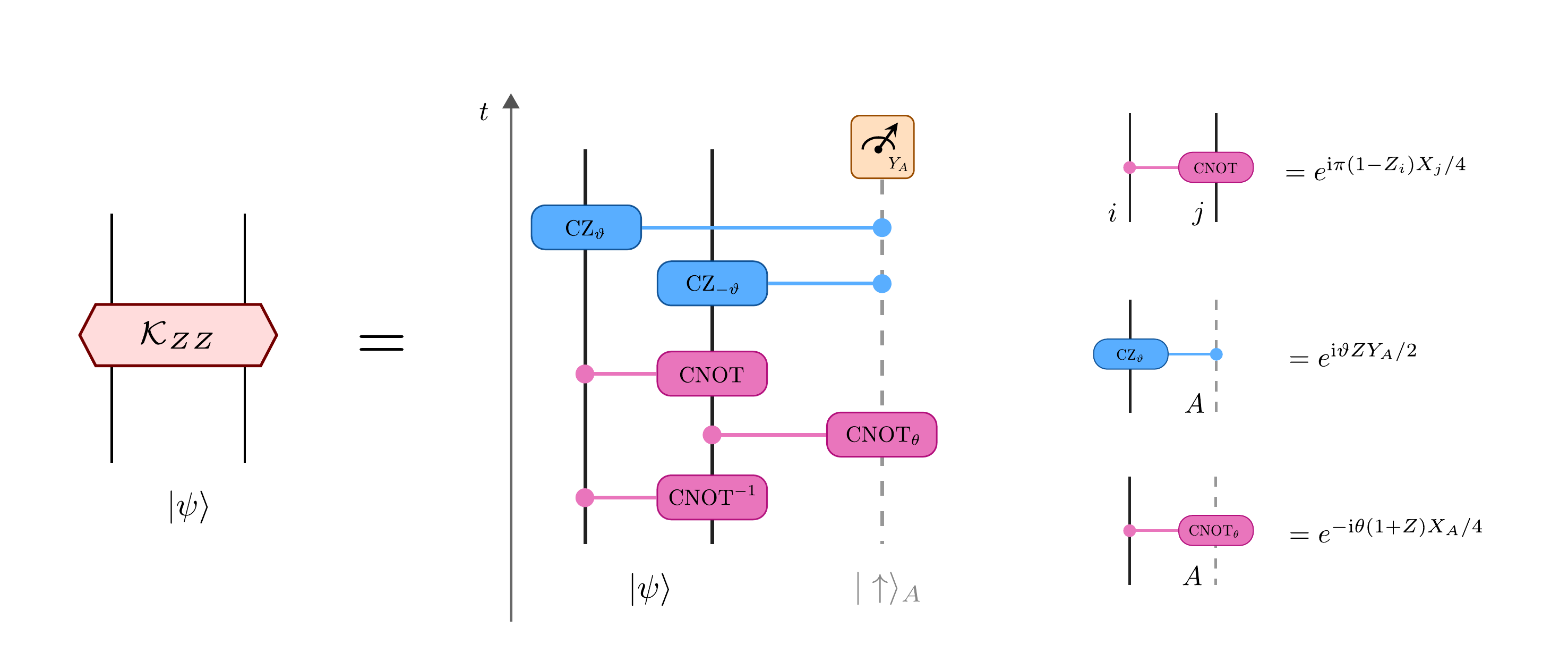}
    \caption{Quantum circuit that realizes the nearest-neighbor Kraus operator.  To match onto the Lindbladian form in the continuous-time limit, we choose $\theta = 2 \arctan \tanh \sqrt{\gamma \delta t}$ and $\vartheta = \sqrt{\gamma \delta t}$.
    At the final time, we measure the ancillary qubit in the $Y_A$ basis.
    The unconditional dynamics are obtained by averaging over measurement outcomes across many trajectories, or by tracing out the ancilla degree of freedom.
    }\label{fig:weak_measurement}
\end{figure*}
The normalized Kraus operator~\eqref{eqn:app_normalized_kraus_operator} can be realized through the circuit in Fig.~\ref{fig:weak_measurement}.
In the first stage, a series of CNOT gates are applied to realize a weak measurement of $Z_i Z_j$ with strength parameterized by $\theta$.
Since the conjugation by the CNOT gate takes $Z_i I_j \to Z_i Z_j$, the set of three CNOT gates may be replaced by a generalized CNOT gate
\begin{equation}
    e^{-\ri \frac{\theta}{2} \frac{1+Z_i Z_j}{2} X_A}
    \,,
\end{equation}
where $X_A$ acts on the ancillary degrees of freedom.
If the ancilla is
initialized in the $\ket{\uparrow}$ state, then the state following the application of the unitary 
can be written in the form
\begin{equation}
    \left(
        \Pi_- \ket{\psi}
        + 
        \cos \frac{\theta}{2}
        \Pi_+
        \ket{\psi}
    \right)
    \ket{\uparrow}_A
    - \ri \sin \frac{\theta}{2}
    \Pi_+ \ket{\psi}
    \ket{\downarrow}_A
    \,,
\end{equation}
where $\Pi_\pm$ is the rank-$2$ projector in the $Z_i Z_j$ basis.
After measuring the ancilla in the $Y_A$ basis, the post-measurement state for measurement outcome $\mu = \pm$
takes the form
\begin{equation}
    \frac{1}{\sqrt{2} }
    \Pi_- \ket{\psi}
    +
    \frac{1}{\sqrt{2} }
    \left(
        \cos \frac{\theta}{2}
        \pm 
        \sin \frac{\theta}{2}
    \right)
    \Pi_+ \ket{\psi}
    \,,
\end{equation}
which matches the real part of the desired Kraus operator~\eqref{eqn:app_normalized_kraus_operator}
\begin{equation}
    \mathrm{Re}(K_\mu) = \frac{1}{\sqrt{2} }
    \left(
        \Pi_-
        + 
        \frac{e^{\mu \sqrt{\gamma \delta t} }}{\sqrt{\cosh 2 \sqrt{\gamma\delta t} } }
        \Pi_+
    \right)
    \,,
\end{equation}
provided we set
\begin{equation}
    \theta = 2 \arctan \tanh \sqrt{\gamma \delta t}  = 2 \sqrt{\gamma \delta
    t} 
    + \mathcal{O}( \delta t^{\frac{3}{2}})
    \,.
\end{equation}
The imaginary part of~\eqref{eqn:app_normalized_kraus_operator} is realized through the second stage of Fig.~\ref{fig:weak_measurement}, which corresponds to a pair of controlled phase rotations conditioned on the measurement outcome.
When the system is in the $Z_i Z_j = +1$ sector, the phase rotations will cancel regardless of the measurement outcome.
On the other hand, when $Z_i Z_j = -1$, then the phases will add constructively for a net rotation $e^{\ri\mu \frac{\vartheta}{2}(Z_j - Z_i)}$, such that the net Kraus operator has the form
\begin{equation}
    K_\mu = \frac{1}{\sqrt{2} }
    \left(
    e^{\ri \mu \frac{\vartheta}{2} ( Z_j - Z_i)}
        \Pi_+
        + 
        \frac{e^{\mu \sqrt{\gamma \delta t} }}{\sqrt{\cosh 2 \sqrt{\gamma\delta t} } }
        \Pi_-
    \right)
    \,,
\end{equation}
which matches exactly the desired form~\eqref{eqn:app_normalized_kraus_operator} with $\vartheta = \sqrt{\gamma \delta t}$.

\section{Mapping to the non-Hermitian Hamiltonian using Weak-Symmetry}\label{sec:app_weak_gauge_sym}
In this section, we analyze the Lindbladian~\eqref{eqn:yl_lindbladian_form} introduced in Section~\ref{sec:operator_dyn_doubled_H} more formally from the perspective of the local weak-symmetry $Z_j \otimes Z_j$.
At the local doubled Hilbert space on each site, it will be useful to redefine the local degrees of freedom in the following form
\begin{equation}
    \begin{split}
        \tau^z = Z \otimes I &\,,\;\;\; \tau^x = X \otimes X
        \\
        \mu^z = Z \otimes Z &\,,\;\;\; \mu^x = I \otimes X \,.
    \end{split}
\end{equation}
The non-Hermitian subspace $\Lambda$ in Eq.~\eqref{eqn:nhh_subspace} corresponds to fixing $\mu^z_j = -1$ on each site $j$.
After doing so, we are left with a local qubit degree of freedom with Pauli operators $\tau^{x,z}$ which we identify with the $\hat{\tau}^{x,z}$ appearing in the non-Hermitian generator of the Yang-Lee transfer matrix $\hat{H}_\mathrm{YL}$.

\subsection{Single-Site Unitary}
The unitary part of the Lindbladian describes single-site $Z$ rotations, whose action takes the form
\begin{equation}
    \ri 
    \frac{
    \theta 
    }{2}
    \sum_j
    Z_j
    \otimes 
    I_j 
    -  
    I_j 
    \otimes 
    Z_j
    = 
    \ri 
    \frac{
    \theta 
    }{2}
    \sum_j
    \tau^z_j
    \left(I_j - \mu^z_j \right)
    = 
    \ri 
    \theta 
    \sum_j
    \tau^z_j
    \,,
\end{equation}
where in the last line have set $\mu^z = -1$.
\subsection{Single-Site \texorpdfstring{$X$}{X}-Dephasing Channel}
Next, there is the single-site non-unitary term of the Lindbladian of the form
\begin{equation}
    \Delta 
    \sum_j 
    X_j \otimes X_j = 
    \Delta     \sum_j 
    \tau^x_j
    \,.
\end{equation}
\subsection{Nearest-Neighbor Channel}
Finally, we consider the nearest-neighbor non-unitary term of the Lindbladian.
After expanding out the exponentials, we obtain
\begin{align}
    \gamma\sum_{\langle i, j\rangle} 
     e^{-\ri \frac{\pi}{4} (Z_i-Z_j)}
    \otimes
    e^{\ri \frac{\pi}{4} (Z_i-Z_j)}
    &=
    \frac{\gamma}{4}\sum_{\langle i, j \rangle}
    \Big[
     (I + \ri Z_j)  \otimes (I-\ri Z_j) 
    + (I + \ri Z_j) \otimes (Z_i Z_j + \ri Z_i) 
     \\
     &
     \;\;\;\;\;\;\;\;\;\;\;\;\;\;\;
     +   (Z_i Z_j - \ri Z_i) \otimes (I - \ri Z_j) + (Z_i Z_j - \ri Z_i) \otimes (Z_i Z_j + \ri Z_i) 
    \Big] \nonumber
    \\
    &=
    \frac{\gamma}{4}\sum_{\langle i, j \rangle}
    \Big[
    I+\mu_j^z + \ri\tau_j^z(I-\mu_j^z)
    +
    \ri\tau_i^z\mu_i^z(I+\mu_j^z) - \tau_i^z\mu_i^z\tau_j^z(I-\mu_j^z)
    \\
    &
    \;\;\;\;\;\;\;\;\;\;\;\;\;\;\;\;\;\;\;\;
    -\ri\tau_i^z(I+\mu_j^z) 
    + \tau_i^z\tau_j^z(I-\mu_j^z)
    +
    \mu_i^z(I+\mu_j^z) + \ri\mu_i^z\tau_j^z(I-\mu_j^z)
    \Big]
    \nonumber
    \\
    &=
    \frac{\gamma}{4}\sum_{\langle i, j \rangle}
    \Big[
    2 \ri \tau^z_j + 2 \tau^z_i \tau^z_j 
    + 2 \tau^z_i \tau^z_j - 2 \ri \tau^z_j
    \Big]
    \\
    &=
    \gamma \sum_{\langle i, j \rangle} \tau^z_i \tau^z_j
    \,.
\end{align}
In the third line, we used $\mu^z_j = \mu^z_i = -1$.
Thus, we conclude that, projected into $\Lambda$, the last term of~\eqref{eqn:yl_lindbladian_form} is equivalent to a nearest-neighbor Ising interaction.

\subsection{Boundary State}
We are also interested in the corresponding boundary state $\bra{\phi_f}, \ket{\phi_i}$ obtained by projecting into $\Lambda$ in Eq.~\eqref{eqn:nhh_projection_dictionary}.

First, for the initial state, we consider $\rho(0) = \ket{+}^N \bra{+}^N$ for simplicity.
Here, $N$ is the system size of the qubit system on which the transfer matrix acts.
Observe that $\rho(0)$ is simultaneously an eigenstate for $\mu^x_j$ and $\tau^x_j$ for all $j$ with eigenvalue $+1$.
Consequently, the state in the doubled Hilbert space can be denoted
\begin{equation}
    |\rho(0)\rangle \rangle = \prod_j \ket{+}_{\mu, j} \ket{+}_{\tau, j} \,,
\end{equation}
where $\ket{\pm}_\mu, \ket{\pm}_\tau$ represent the $\mu^x = \pm 1$ and $\tau^x = \pm 1$ states, respectively.
Similarly, for the initial state $\rho(0) = \ket{-}^N \bra{-}^N$, the initial state can be written as
\begin{equation}
    |\rho(0)\rangle \rangle = \prod_j \ket{-}_{\mu, j} \ket{+}_{\tau, j} \,.
\end{equation}

On the other hand, evaluating the trace of a state can be represented as an overlap between states in the doubled Hilbert space
\begin{equation}
    \Tr \rho = \langle \langle I | \rho \rangle \rangle
    \,,
\end{equation}
where $\langle \langle I |$ is a Bell state between the two copies of the Hilbert space
\begin{equation}
    \langle \langle
    I | = 
    \prod_j
    \left(
    \sum_{\pm}
    \bra{\pm}_j
    \otimes
    \bra{\pm}_j
    \right)
    =
    2^{N/2} \prod_j \bra{\uparrow}_{\mu, j} \bra{+}_{\tau, j}
    \,.
\end{equation}
Here, $\bra{\uparrow}_\mu, \bra{\downarrow}_\mu$ are the $+1$ and $-1$ eigenstates of $\mu^z$.

Finally, we will be interested in the expectation values of observables $O$ evaluated with respect to the state $\rho$
\begin{equation}
    \Tr O \rho = \langle \langle I | O \otimes I |\rho\rangle\rangle \,.
\end{equation}
The operator $O$ acts to the left, and can be absorbed into final boundary state by its action $\langle \langle I | O = \langle \langle O |$.
For simplicity, we focus the case of $O = \prod_j X_j$.  Thus we have
\begin{equation}
    \langle\langle \prod_j X_j | = 2^{N/2} \prod_j \bra{\uparrow}_{\mu, j} \bra{+}_{\tau, j}
    \mu_j^x \tau_j^x = 
    2^{N/2} \prod_j \bra{\downarrow}_{\mu, j} \bra{+}_{\tau, j}
    \,.
\end{equation}
Similarly, the operator $O = \prod_j (\ri Y_j)$ yields the boundary state
\begin{equation}
    \langle\langle \prod_j (\ri Y_j) | = 2^{N/2} \prod_j \bra{\uparrow}_{\mu, j} \bra{+}_{\tau, j}
    \tau^z_j 
    \mu_j^x 
    \tau_j^x = 
    2^{N/2} (-1)^N \prod_j \bra{\downarrow}_{\mu, j} \bra{-}_{\tau, j}
    \,.
\end{equation}

Thus, the expectation value of $\prod_j X_j$ in the state obtained by evolving $\ket{+}^N \bra{+}^N$ under the Lindbladian for time $T$ maps to the Yang-Lee partition function with free boundary conditions
\begin{equation}
    \Tr \left(\prod_j X_j e^{T \mathcal{L}}[\rho(0)] \right)= 
    \langle \langle \prod_j X_j| e^{T\mathcal{L}}\ket{\rho(0)}
    = \bra{+}^N e^{- T \hat{H}_\mathrm{YL}} \ket{+}^N 
    = \mathcal{Z}^\mathrm{free}_\mathrm{YL}
    \,.
\end{equation}
\subsection{\texorpdfstring{$\mathcal{PT}$}{PT}-Symmetry Operator}\label{app:pt_symmetry}
The anti-unitary $\mathbb{Z}_2$ Hermiticity-preserving symmetry of the Lindbladian takes the form of complex conjugation $\mathcal{K}$ followed by a SWAP between the forward and backwards Hilbert spaces.

In the qubit degrees of freedom, this can be written
\begin{align}
    \mathcal{K} \prod_j \mathrm{SWAP}_j  &=
    \mathcal{K} \prod_j \frac{I_j \otimes I_j + X_j \otimes X_j + Y_j \otimes Y_j + Z_j \otimes Z_j}{2}
    \\
    &=
    \mathcal{K} 
    \prod_j 
    \frac{
    1 + \tau^x_j - \tau^x_j \mu^z_j +\mu^z_j
    }
    {
    2}
    \\
    &= \mathcal{K}
    \prod_j
    \tau^x_j
    \,,
\end{align}
where in the last line we fixed the weak-symmetry sector $\mu^z_j = -1$.
Thus, we conclude that the $\mathcal{PT}$-symmetry operator $\mathcal{K} \prod_j \hat{\tau}^x_j$ of $\hat{H}_\mathrm{YL}$ is exactly descended from the Hermiticity-preserving symmetry of the underlying quantum channel.
\end{document}